%% file: iclr2027_conference.tex
\documentclass{article} 
\usepackage{iclr2027_conference,times}

\input{math_commands.tex}

\usepackage{hyperref}
\usepackage{url}
\usepackage[table]{xcolor}
\usepackage{hyperref}
\usepackage{url}
\usepackage{multirow}
\usepackage{booktabs}
\usepackage{graphicx}
\usepackage{booktabs}
\usepackage{multirow}
\usepackage[table]{xcolor}
\usepackage{threeparttable}
\usepackage{subcaption}
\usepackage{wrapfig}
\usepackage[table]{xcolor}
\definecolor{ablationyellow}{RGB}{255,249,220}
\usepackage{tabularx}
\usepackage{booktabs}
\usepackage{array}
\usepackage{longtable}

\author{
  \textbf{Suifeng Zhao\textsuperscript{1}},
  \textbf{Zida Liu\textsuperscript{1}},
  \textbf{Lei Sun\textsuperscript{2}},
  \textbf{Xun Wang},
  \textbf{Daojian Zeng\textsuperscript{3}},
  \textbf{Jun Gao\textsuperscript{1*}},
  \textbf{Sujian Li\textsuperscript{4}}\thanks{Corresponding authors.}
  \\
  \textsuperscript{1}Key Laboratory of High Confidence Software Technologies,
  CS, Peking University, China \\
  \textsuperscript{2}Panasonic Connect Co., Ltd., Tokyo, Japan  \textsuperscript{3}Hunan Normal University, China \\
  \textsuperscript{4}Key Laboratory of Computational Linguistics(MOE), CS, Peking University, China \\
  \texttt{{sfzhao25}@stu.pku.edu.cn}, \texttt{\{gaojun,lisujian\}@pku.edu.cn}
}

\title{AttriSwarm: Orchestrating Agent Swarms for Cross-Page Visual Attribution}

\iclrfinalcopy

\begin{document}
\maketitle
\fancyhead[L]{Preprint}

\maketitle

\begin{abstract}
Multimodal large language models (MLLMs) increasingly address knowledge-intensive tasks by integrating evidence scattered across multiple document pages into a final answer. Because such answers are synthesized from many pages, the link between the answer and its supporting visual evidence can become obscured, hindering verification and trust. Cross-page visual attribution is therefore essential: it requires identifying the supporting pages and localizing fine-grained evidence regions within them, thereby making multi-page reasoning transparent and auditable. Existing approaches either process multiple pages jointly or iteratively refine evidence within a single trajectory, risking cross-page context pollution and, for interactive methods, substantial sequential overhead. We introduce \textbf{AttriSwarm}, an agent-swarm framework that assigns these concerns to separate roles: a leader handles global cross-page reasoning, while subagents perform page-wise attribution in isolation and in parallel. To support training and evaluation, we construct \textbf{AttriDoc}, a cross-page QA dataset linking supporting pages to their annotated evidence regions, and use it for role-specific training of AttriSwarm, improving answer accuracy, page-level $F_1$, and region-level IoU over its training-free counterpart by an average of 18.94, 29.34, and 23.04 percentage points, respectively, across three open-source backbones and three datasets.
\end{abstract}

\begin{figure*}[h]
  \centering
\includegraphics[width=0.8\linewidth]{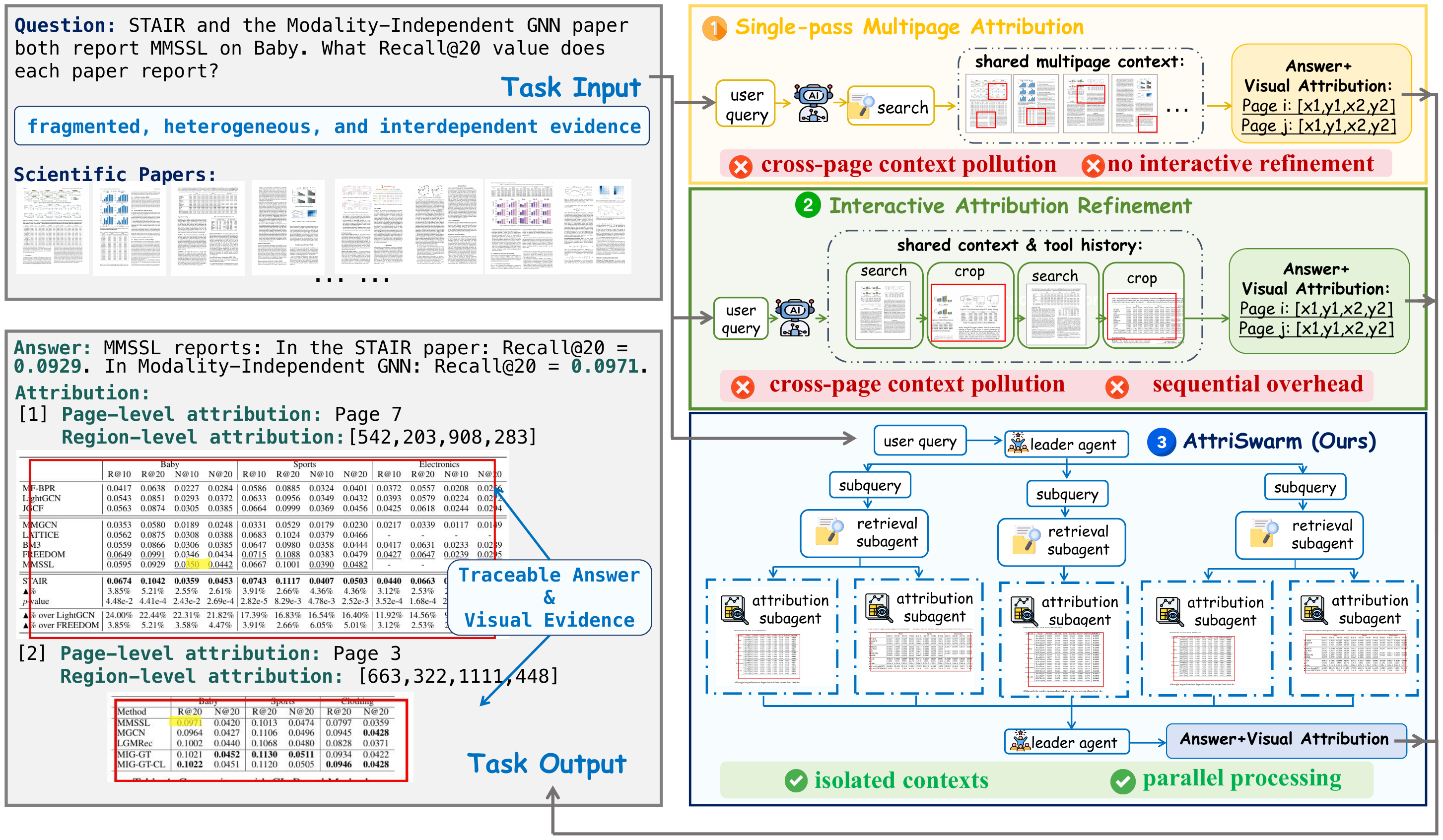}
  \caption{\textbf{Example and paradigms for cross-page visual attribution.}
\emph{Single-pass multipage attribution} jointly processes candidate pages without interactive refinement.
\emph{Interactive attribution refinement} supports iterative evidence localization but introduces sequential overhead.
Both paradigms operate in shared contexts and are susceptible to cross-page context pollution.
\textbf{AttriSwarm} isolates and parallelizes page-wise attribution while preserving global evidence integration.}
  \label{fig:paradigm}
  \vspace{-1\baselineskip}
\end{figure*}
\section{Introduction}\label{sec:intro}
\begin{wrapfigure}{r}{0.45\textwidth}
    \vspace{-0.8\baselineskip}
    \centering
    \includegraphics[
        width=0.85\linewidth,
        page=1
    ]{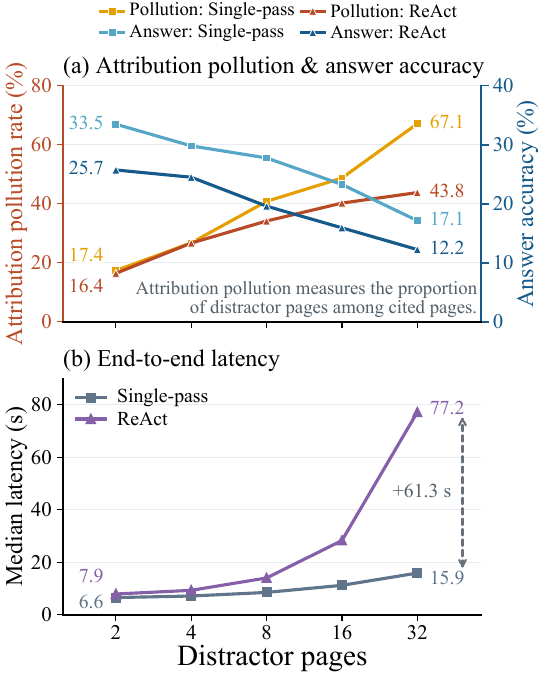}
    \caption{\textbf{Cross-page context pollution and sequential overhead.}
    As distractor pages increase, (a) both methods exhibit higher attribution pollution and lower answer accuracy; (b) ReAct exhibits steeper latency growth.}
    \label{fig:motivation_pollution_latency}
    \vspace{-1.4\baselineskip}
\end{wrapfigure}

Recent advances in multimodal large language models (MLLMs) have enabled their application to complex knowledge-intensive tasks, such as multimodal deep research~\citep{DBLP:journals/corr/abs-2601-22060,DBLP:journals/corr/abs-2601-12346,DBLP:journals/corr/abs-2601-09278,DBLP:journals/corr/abs-2603-01050,zeng2026visiondeepresearchbenchmarkrethinkingvisual,huang2026visiondeepresearchincentivizingdeepresearchcapability,yang2025multimodaldeepresearchergeneratingtextchart} and scientific literature analysis~\citep{DBLP:conf/acl/XiongYLCD26,DBLP:journals/corr/abs-2505-19897,li2026justreasonjustscan}. When the evidence supporting an answer is contained within a single page, visual attribution can localize the relevant regions within that page, making the answer traceable to its source. However, many knowledge-intensive tasks require models to search, select, and integrate evidence distributed across multiple pages. In such scenarios, \textbf{cross-page visual attribution} requires identifying both the supporting pages and the corresponding evidence regions within each page, allowing generated answers to be verified against their source material, as illustrated in Figure~\ref{fig:paradigm}. Compared with the single-page setting, cross-page attribution is more challenging because the evidence required for an answer is often fragmented across pages, heterogeneous in form, and is highly interdependent.

Existing approaches to cross-page visual attribution broadly follow two paradigms. \emph{Single-pass multipage attribution} processes all candidate pages jointly~\citep{DBLP:conf/acl/MaZKZCL25,DBLP:conf/emnlp/ZhaoJLG25}, but cannot revisit ambiguous evidence or inspect finer-grained details through subsequent interactions. To address this limitation, \emph{interactive attribution refinement} supports iterative searching, cropping, or zooming~\citep{DBLP:conf/nips/WangDZCCWXHZ25,DBLP:journals/corr/abs-2602-24134,DBLP:journals/corr/abs-2604-14967,zhao2026mciterlreliablemultimodalrag}. However, despite this additional flexibility, both paradigms can suffer from \textbf{cross-page context pollution}~\citep{phillips2026building}, where irrelevant pages in a shared context interfere with evidence selection and localization. To assess their susceptibility to this problem, we inject 2--32 distractor pages into a subset of CiteVQA~\citep{ma2026citevqa} and compare single-pass attribution with a representative interactive refinement method, ReAct~\citep{yao2023reactsynergizingreasoningacting}, under the same Qwen3-VL-4B-Instruct backbone~\citep{qwen3technicalreport}. As shown in Figure~\ref{fig:motivation_pollution_latency} (a), both methods exhibit higher attribution pollution (the proportion of distractor pages among cited pages) and lower answer accuracy as distractors increase. Meanwhile, ReAct's latency increases sharply with page count (Figure~\ref{fig:motivation_pollution_latency} (b)), highlighting the \textbf{sequential overhead} of serial page-wise attribution.

Architecturally, these issues arise from coupling heterogeneous reasoning states and page-wise operations within a single trajectory. While recent studies on agent swarms and multi-agent delegation leverage context isolation, task decomposition, and parallel execution to alleviate this burden~\citep{DBLP:journals/corr/abs-2602-04634,DBLP:journals/corr/abs-2607-08662,DBLP:journals/corr/abs-2602-02276,zhang2026orchestrao1omnimodalagentorchestration,virk2026swarmresearchorchestratingcodingagents}, cross-page visual attribution faces a unique challenge. Its primary bottleneck is the intrinsic conflict between isolating local relevance assessment and region localization, and preserving the global evidence relationships essential for reasoning.

Building on these capabilities, we introduce \textbf{AttriSwarm}, an agent-swarm framework that preserves global evidence integration while isolating local attribution. The leader decomposes the query into subqueries that guide parallel retrieval and specify the evidence sought in local attribution. Each attribution subagent receives a single candidate page and its corresponding subquery to assess relevance and localize supporting regions within an isolated context. Attribution runs in parallel to reduce sequential overhead, while the leader integrates the returned evidence for cross-page reasoning and generates the final answer with visual attributions.

To support training and evaluation, we construct \textbf{AttriDoc} with cross-page QA and page- and region-level annotations: \textbf{AttriDoc-Paper} is synthesized from tables linked by shared entities across scientific papers, while \textbf{AttriDoc-Slide} augments multipage QA pairs from SlideVQA~\citep{DBLP:conf/aaai/TanakaNNHSS23} with region-level attribution annotations. To make this decomposition effective with general-purpose MLLMs, we further develop a role-specific training strategy for the leader and attribution subagents, improving answer accuracy, page-level $F_1$, and region-level IoU over its training-free counterpart by an average of 18.94, 29.34, and 23.04 percentage points, respectively, across three open-source backbones and three datasets.

Our contributions are threefold for complex cross-page visual attribution: (1) Architecturally, we propose \textbf{AttriSwarm}, an agent-swarm framework that isolates and parallelizes page-wise attribution while preserving global cross-page reasoning. (2) We construct \textbf{AttriDoc}, a multimodal dataset comprising scientific-paper and slides, with cross-page QA pairs and page- and region-level attributions. (3) We develop a role-specific training strategy for leader and attribution subagents, and extensive experiments demonstrate improvements in both answer accuracy and visual attribution.

\section{Architecture of AttriSwarm}~\label{sec:attriswarm}
\begin{figure*}[t]
  \centering
  \includegraphics[width=\linewidth]{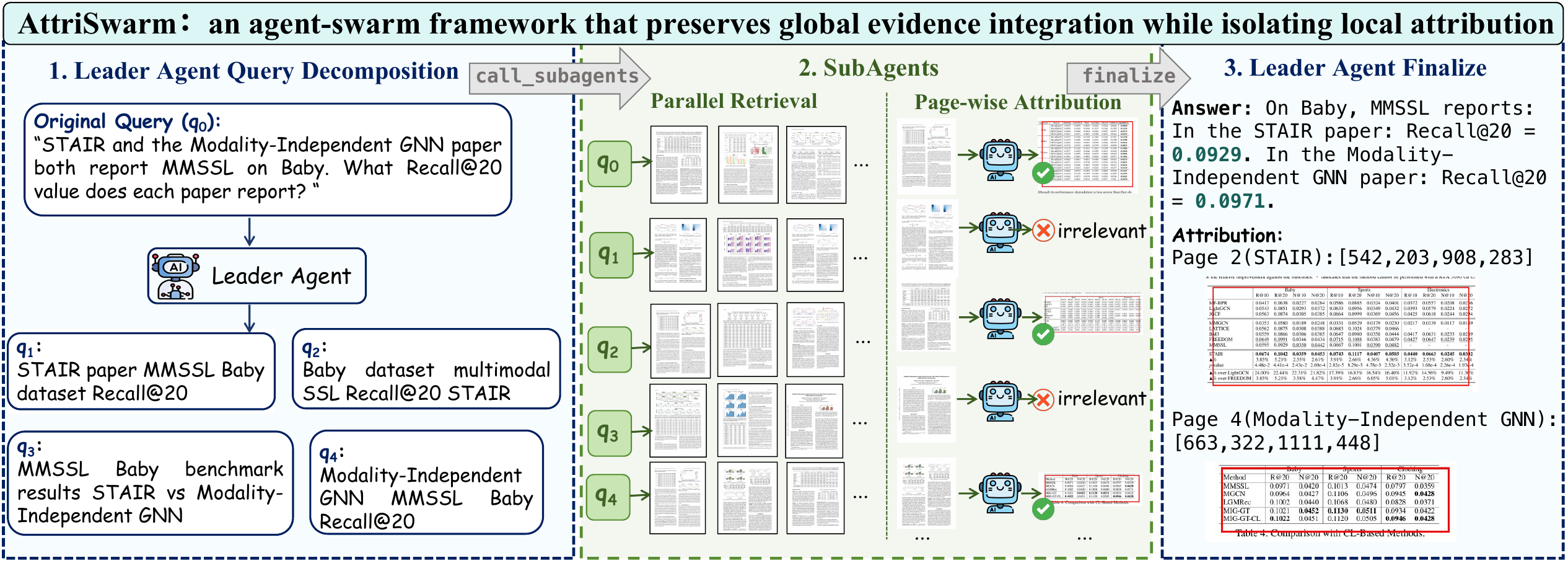}
  \caption{\textbf{Overview of AttriSwarm.} AttriSwarm decomposes cross-page reasoning into leader-based query decomposition, parallel retrieval and page-wise attribution of subagents, and final evidence aggregation by the leader agent.}
   \vspace{-1\baselineskip}
  \label{fig:AttriSwarm}
\end{figure*}

We present AttriSwarm (Figure~\ref{fig:AttriSwarm}), a swarm-based framework that mitigates cross-page context pollution and sequential overhead. The leader decomposes the query and integrates evidence, while parallel, context-isolated subagents assess page relevance and localize supporting regions. 

\subsection{Task Formulation}
We formulate the cross-page visual attribution task in a retrieval-augmented multimodal setting. 
Given a user query $q$ and a document page corpus $\mathcal{D} = \{p_1, p_2, \ldots, p_{|\mathcal{D}|}\}$, the goal is to generate a grounded answer $a$ along with explicit visual attributions at both the page and region levels, denoted as $\mathcal{S}_q$ and $\mathcal{B}_q$, respectively:
\begin{equation}
    (q, \mathcal{D}) \rightarrow (a, \mathcal{S}_q, \mathcal{B}_q)
\end{equation}

Given a query $q$, a retrieval module retrieves the top-$K$ candidate pages $\mathcal{C}_q=\{p_k\}_{k=1}^K\subset\mathcal{D}$. The model then identifies the supporting pages $\mathcal{S}_q\subseteq\mathcal{C}_q$, defining the \textbf{page-level attribution}, and localizes evidence regions on each $p\in\mathcal{S}_q$ as $B_p=\{b_{p,j}\}_{j=1}^{m_p}$, where $b_{p,j}=(x_1,y_1,x_2,y_2)$. The \textbf{region-level attribution} is therefore $\mathcal{B}_q=\{(p,B_p)\mid p\in\mathcal{S}_q\}$.

\subsection{Leader Agent for Global Cross-Page Reasoning}
The leader focuses on what evidence is needed and whether it jointly supports an answer, leaving retrieval and page-wise attribution execution to subagents. It uses two actions: \texttt{call\_subagents}, which triggers retrieval and subsequent page-wise attribution based on subqueries, and \texttt{finalize}, which terminates evidence acquisition and produces the final answer with visual attributions.
\paragraph{Query Decomposition and Subagent Invocation.}
Given a user query $q$, the leader first analyzes the required evidence and reformulates it into a set of retrieval-oriented subqueries, $Q_q=\{q_0,q_1,\ldots,q_M\}$,
where $q_0=q$ preserves the original query and the remaining subqueries provide complementary retrieval views of the required evidence. As illustrated in Figure~\ref{fig:AttriSwarm}, the leader decomposes the original query into multiple subqueries targeting complementary aspects and executes \texttt{call\_subagents}, which triggers parallel retrieval followed by page-wise attribution. In this way, the leader specifies what evidence should be acquired, while retrieval and visual attribution are delegated to specialized subagents and executed outside the leader's reasoning context. We describe these two actions in Section~\ref{sec:subagent}.

\paragraph{Cross-Page Attribution Selection and Answer Synthesis.}
After the subagent calls are completed, the leader receives the relevant pages together with their localized supporting regions and assesses whether the collected evidence is sufficient to answer the query. If so, it executes \texttt{finalize}, which aggregates and verifies the page-specific evidence to produce the final answer $a$, along with the selected supporting pages $\mathcal{S}_q$ and their corresponding region-level attributions $\mathcal{B}_q$.

\subsection{Parallel Subagents for Retrieval and Page-Wise Attribution}~\label{sec:subagent}
The subagent layer organizes retrieval by subquery and attribution by page, allowing evidence acquisition to proceed in parallel while keeping local attribution contexts isolated.

\paragraph{Parallel Retrieval.}
Given the retrieval-oriented subqueries generated by the leader, each retrieval subagent searches the document corpus for candidate pages relevant to its assigned subquery. As these retrieval tasks target complementary evidence and can be executed independently, retrieval proceeds in parallel rather than through a sequential single-agent trajectory. The retrieved pages are then collected and deduplicated by the externalized harness, with their associated subqueries preserved for subsequent attribution.

\paragraph{Isolated Page-wise Attribution.}
Each attribution subagent receives a single candidate page $p_i$ and its corresponding subquery $\tilde{q}_i\in Q_q$, assesses the page's relevance to $\tilde{q}_i$, and localizes supporting regions when the page is relevant.
Formally, $(\tilde{q}_i,p_i)\rightarrow(r_i,B_i)$, where $r_i\in\{0,1\}$ indicates page relevance to $\tilde{q}_i$, and $B_i=\{b_{i,1},\ldots,b_{i,m_i}\}$ contains the localized evidence boxes when $r_i=1$; otherwise, $B_i=\emptyset$.
The externalized harness collects these outputs and forwards only pages with $r_i=1$, along with their localized regions, to the leader for cross-page reasoning and synthesis.
By operating on a single candidate page within an isolated context, each subagent limits cross-page context pollution while enabling parallel attribution across multiple pages.
\subsection{Externalized Harness for State and Evidence Management}
To support the coordination between the leader and parallel subagents, AttriSwarm employs an externalized harness as a programmatic state-management layer. It maintains four types of externalized state: (1) a \texttt{Query Registry} that stores the original query and its decomposed subqueries, (2) a \texttt{Retrieved Page Pool} that collects and deduplicates retrieved pages while preserving their retrieval scores and provenance information, (3) a \texttt{Dispatch Table} that tracks the assignment of candidate pages to attribution subagents, and (4) an \texttt{Evidence Ledger} that records page relevance assessments and localized supporting regions returned by the subagents. By externalizing these workflow states, the harness keeps intermediate retrieval and attribution information outside the agents' reasoning contexts, ensuring both the leader and each subagent access only the information required for its current stage, while preserving explicit evidence throughout the workflow.

\section{AttriDoc Construction}~\label{sec:attridoc}
Cross-page visual attribution requires page- and region-level evidence annotations. \textbf{AttriDoc} comprises 5,471 training and 1,236 test QA pairs across \textbf{AttriDoc-Paper} and \textbf{AttriDoc-Slide}, annotated with supporting pages and evidence boxes. 
\subsection{Data Sources and Annotation}

\paragraph{AttriDoc-Paper.}
This subset is constructed from computer science papers for cross-page reasoning over experimental results. We collect seed papers from arXiv and their cited papers, parsing PDF and HTML representations to extract experimental tables. GPT-5.1~\citep{singh2026openaigpt5card} groups evidence across papers by shared entities (methods, datasets, and metrics) and synthesizes QA pairs that require cross-page evidence integration from related tables and page images. We extract table boxes using MinerU~\citep{wang2026mineru2}, where source pages and boxes serve as page- and region-level attributions, respectively.

\paragraph{AttriDoc-Slide.}
This subset is built from the multipage QA pairs in SlideVQA~\citep{DBLP:conf/aaai/TanakaNNHSS23}. We retain the original questions, answers, and supporting-pages and augment them with region-level bounding boxes. Following AttriSwarm, we use Gemini-3-Flash-Preview~\citep{DBLP:journals/corr/abs-2507-06261} as the attribution subagent to localize supporting regions in trajectory generation.

\subsection{Quality Control and Statistics}

\paragraph{Quality Control.}
\input{tables/data1}
For AttriDoc-Paper, we retain QA pairs only when GPT-5.1~\citep{singh2026openaigpt5card}, Gemini-3-Flash-Preview~\citep{DBLP:journals/corr/abs-2507-06261}, and Claude Sonnet 4.6~\citep{anthropic2024claude35sonnet} unanimously agree that the answers are correct based on the questions and cropped evidence images. We further manually filter the test set for: (1) clearly specified information requests; (2) unambiguous references to methods, datasets, and metrics; (3) necessary and sufficient evidence; and (4) answer values that are consistent with the corresponding tables. For AttriDoc-Slide, GPT-5.1 evaluates whether the cropped evidence supports the answer, covers the required information, and avoids excessive irrelevant content. Failed annotations undergo up to three rounds of revision, and cases that still fail are discarded.

\paragraph{Dataset Statistics.} Table~\ref{tab:attridoc-statistics} summarizes the dataset scale and the distribution of distinct annotated evidence pages per test question. After quality control, AttriDoc-Paper contains 4,079 training and 669 test QA pairs, with a retrieval corpus of 42,306 pages from 2,359 papers. AttriDoc-Slide contains 1,392 training and 567 test QA pairs, with a retrieval corpus of 19,749 slide images. 

\section{Role-Specific Training}~\label{sec:train}
While AttriSwarm provides the architectural decomposition needed to isolate and parallelize page-wise attribution, general-purpose MLLMs are not inherently optimized for the specialized roles introduced by this decomposition. In particular, false-negative relevance decisions may discard useful pages, while inaccurate localization degrades the evidence returned to the leader. We therefore develop role-specific training for both the leader and attribution subagents.

\subsection{Trajectory-Based Supervised Fine-Tuning}
We perform SFT separately for the leader and attribution subagents using role-specific interaction trajectories. The trajectories are constructed as follows. For both AttriDoc subsets, GPT-5.1 serves as the leader. It generates retrieval-oriented subqueries and invokes subagents to retrieve candidate pages. To avoid redundant retrieval, we discard any subquery whose retrieved pages are fully covered by other retained queries. Each retrieved page is then paired with the first retained query that retrieves it, with the original user query ordered first. Page-wise attribution responses are derived from AttriDoc-Paper annotations, or generated by Gemini-3-Flash-Preview for AttriDoc-Slide. The leader then integrates the returned evidence and invokes finalize to produce an answer with selected visual attributions. These trajectories provide role-specific supervision. Leader trajectories supervise query decomposition, subagent invocation, evidence aggregation, and final response generation. Attribution trajectories supervise page relevance assessment and supporting-region localization.

\subsection{Attribution-Aware Reinforcement Learning}
After SFT, both roles are optimized using Group Relative Policy Optimization (GRPO)~\citep{DBLP:journals/corr/abs-2402-03300}, with separate rewards aligned with their responsibilities.

\paragraph{Attribution Subagent Optimization.}
We jointly optimize both page-level relevance assessment and supporting-region localization via a unified reward:
\begin{equation}
R_{\mathrm{sub}}
=
\lambda_r R_{\mathrm{region}}
+
\lambda_{\mathrm{rel}} R_{\mathrm{rel}}
+
\lambda_f R_{\mathrm{fmt}},
\end{equation}
where $R_{\mathrm{region}}$ uses $\mathrm{IoU}=|B_p\cap B_g|/|B_p\cup B_g|$, with $B_p$ and $B_g$ denoting the geometric unions of predicted and reference evidence boxes on a page, respectively.
$R_{\mathrm{rel}}$ measures the correctness of page relevance decisions, penalizing missed evidence on relevant pages and spurious predictions on irrelevant pages.
$R_{\mathrm{fmt}}$ penalizes outputs violating the required attribution format.

\paragraph{Leader Agent Optimization.}
The leader is optimized for answer correctness, page-level attribution, and output validity:
\begin{equation}
R_{\mathrm{leader}}
=
\mathbb{I}_{\mathrm{valid}}
\left(
\lambda_a R_{\mathrm{ans}}
+\lambda_p R_{\mathrm{page}}
+\lambda_{ap}R_{\mathrm{ans}}R_{\mathrm{page}}
\right),
\end{equation}
where $R_{\mathrm{ans}}$ evaluates answer correctness via an LLM-based verifier, and $R_{\mathrm{page}}$ is the $F_1$ score
between the predicted supporting pages and the retrieved
reference pages, scaled by the fraction of all reference
pages successfully retrieved, and $\mathbb{I}_{\mathrm{valid}}$ indicates whether the final output satisfies the required structured format. 

\section{Experiments and Analysis}~\label{sec:experimental_setup}
We evaluate on AttriDoc-Paper and AttriDoc-Slide as in domain settings, and on 1,175 out of domain CiteVQA questions~\citep{ma2026citevqa}. The CiteVQA subset contains questions with accessible source PDFs and is shared across all methods.
\paragraph{Evaluation Metrics.}
We evaluate answer quality and evidence attribution using four metrics: (1) Answer Accuracy (Acc.) measures final answer correctness with an LLM judge; (2) Page Attribution $F_1$ (Page $F_1$) evaluates whether the correct supporting pages are identified; (3) Region Attribution IoU (Reg. IoU) measures fine-grained evidence localization, averaged over all ground-truth supporting pages so that missed pages are also penalized; and (4) Joint Attribution Score (JAS) combines page selection and region grounding:
\begin{equation}
\mathrm{JAS}
= \frac{1}{N}\sum_{q=1}^{N}
F_{1,q}\cdot\mathrm{IoU}_q.
\end{equation}
Here, $N$ is the number of questions, and $F_{1,q}$ and $\mathrm{IoU}_q$ are the page and region attribution scores for question $q$. All metrics are averaged across questions and reported as percentages. Together, these metrics measure answer correctness and coarse-to-fine attribution quality. 

\paragraph{Baselines.}
We compare AttriSwarm with two attribution paradigms (Section~\ref{sec:intro}): (1) Single-pass jointly processes the top-$K$ retrieved pages ($K \in \{10,20\}$) in a shared multimodal context, producing the answer and page-indexed evidence boxes in one call; (2) Interactive attribution includes training-free ReAct, which alternates reasoning and sequential search within a single trajectory, with final outputs adapted to include the answer and page-indexed evidence boxes. We also evaluate released checkpoints of VRAG-RL~\citep{DBLP:conf/nips/WangDZCCWXHZ25}, UniDoc-RL~\citep{DBLP:journals/corr/abs-2604-14967}, and MCite-RL~\citep{zhao2026mciterlreliablemultimodalrag}, preserving their search and crop protocols. Single-pass and ReAct use Qwen3-VL-4B-Instruct, Qwen3-VL-8B-Instruct, Qwen2.5-VL-7B-Instruct, GPT-5.1, and Claude Sonnet 4.6. We also include training-free AttriSwarm with the same architecture and no task-specific training (details in Appendix~\ref{app:implementation}).
\subsection{Main Results}
\paragraph{Overall Performance.}
\input{tables/main_experiment}

As shown in Table~\ref{tab:main_results}, AttriSwarm improves both answer correctness and evidence attribution. For example, with Qwen3-VL-8B-Instruct on AttriDoc-Slide, it raises answer accuracy from 57.14\% to 68.43\% and JAS from 6.51\% to 45.86\% over the strongest single-pass baseline. The gains in Page $F_1$ and Reg. IoU further indicate stronger coarse-to-fine evidence attribution and better consistency between page selection and region attribution. Overall, these results demonstrate that AttriSwarm improves both cross-page reasoning and end-to-end attribution.

\paragraph{Effectiveness of Role-Specific Training.}
Compared with its training-free counterpart, trained AttriSwarm consistently improves all four metrics across the nine open-source backbone--dataset combinations. Averaged over these combinations, role-specific training improves answer accuracy and JAS by 18.94 and 20.63 percentage points, respectively. Improvements on out-of-domain CiteVQA further suggest that the learned attribution capabilities transfer beyond the training domains.
\subsection{Ablation Study}
\input{tables/ablation.tex}
To disentangle the contributions of training each agent role, we independently vary whether each agent role is trained (Table~\ref{tab:role-training-ablation-all}). Training attribution agents alone improves all four metrics across all settings. For example, on AttriDoc-Paper with Qwen3-VL-4B-Instruct, it raises page-level $F_1$ from 14.45\% to 60.33\% and region-level IoU from 7.62\% to 49.31\%. Leader training yields limited or inconsistent attribution gains on its own, but improves answer accuracy across all nine backbone--dataset settings when the attribution agents are trained. These results suggest complementary benefits: attribution agent training strengthens evidence grounding, while leader training helps translate that evidence into more accurate answers. We also evaluate different model assignments for the leader and attribution agents in Appendix~\ref{app:model_allocation}.

\subsection{Further Analysis}

\paragraph{Robustness to Cross-page Context Pollution.}
\begin{figure}[t]
    \vspace{-6pt}
    \centering
    \includegraphics[
        width=\linewidth,
        page=1
    ]{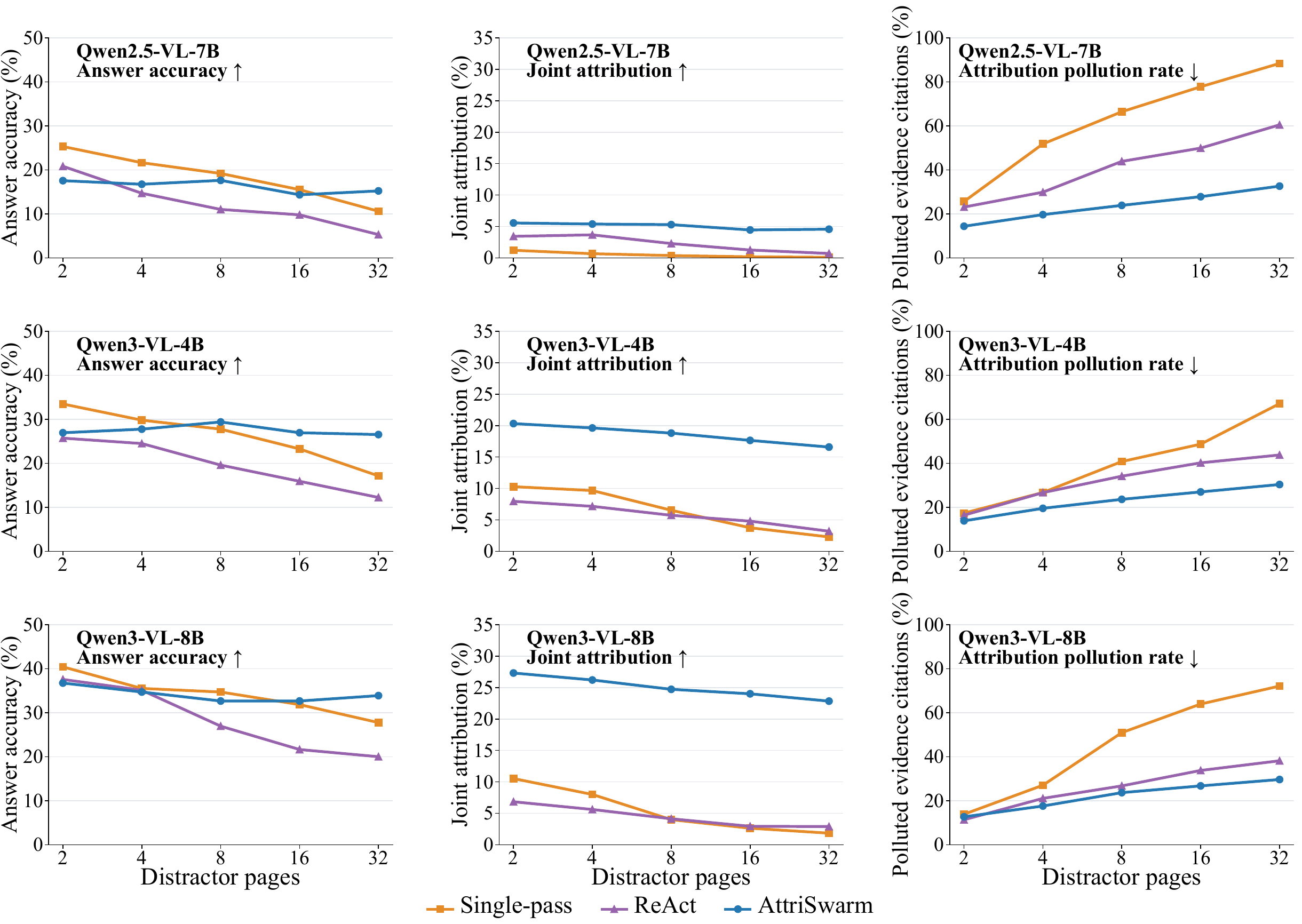}
    \caption{Robustness to increasing numbers of distractor pages
    on CiteVQA across three backbones.}
    \label{fig:context-pollution}
    \vspace{-14pt}
\end{figure}
Building on the diagnostic in the introduction, we add 2--32 retrieved distractor pages to 245 CiteVQA questions, retaining all annotated evidence pages and matching page order across methods. We measure accuracy, JAS, and attribution pollution (the mean fraction of distractors among mapped cited pages), excluding empty outputs and error cases. Across three backbones, the mean accuracy drop over this range is 1.86 percentage points for AttriSwarm versus 14.56 and 15.51 for Single-pass and ReAct, respectively (Figure~\ref{fig:context-pollution}).
AttriSwarm's attribution pollution grows more slowly; despite declining with noise, its JAS remains highest at 32 distractors across all three backbones. These results suggest that context isolation and evidence filtering before global aggregation help mitigate cross-page context pollution.

\paragraph{Inference Efficiency.}
\begin{wrapfigure}{r}{0.4\textwidth} 
    \centering
    \includegraphics[
        width=\linewidth, 
        page=1
    ]{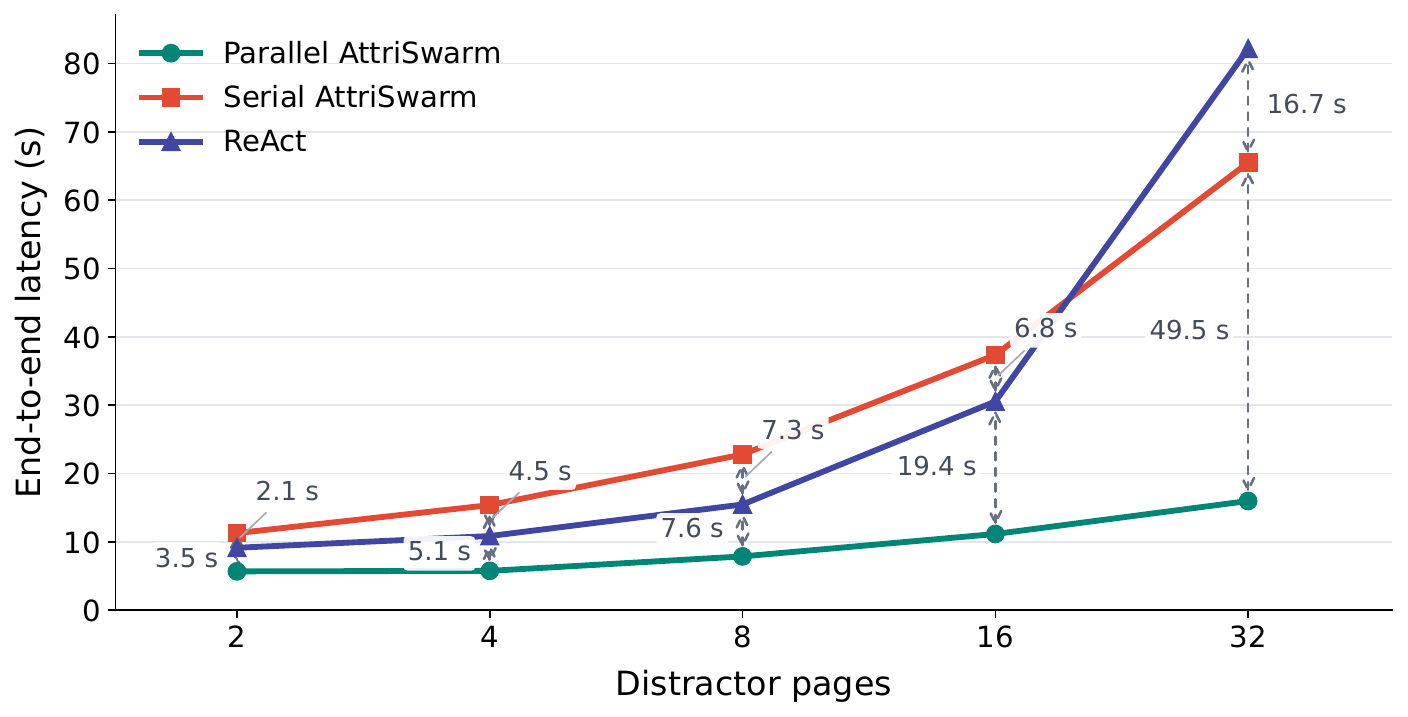} 
    \caption{Efficiency comparison.}
    \label{fig:efficiency-three-methods}
    \vspace{-0.8\baselineskip} 
\end{wrapfigure}

To examine whether parallel attribution reduces the sequential overhead identified in the introduction, we compare parallel AttriSwarm, its serial variant, and ReAct on the same frozen CiteVQA page sets using Qwen3-VL-4B-Instruct, measuring mean end-to-end latency over successful runs. As shown in Figure~\ref{fig:efficiency-three-methods}, at 32 distractor pages, parallel AttriSwarm takes 15.99 seconds, achieving approximately $4.1\times$ and $5.1\times$ speedups over its serial variant and ReAct, respectively. The comparison with the serial variant highlights the benefit of parallel attribution in reducing sequential overhead.

\paragraph{Answer-Supporting Attribution.}
\begin{wrapfigure}[8]{r}{0.4\textwidth}
    \vspace{-3.5\baselineskip}
    \centering
    \includegraphics[
        width=\linewidth,
        page=1
    ]{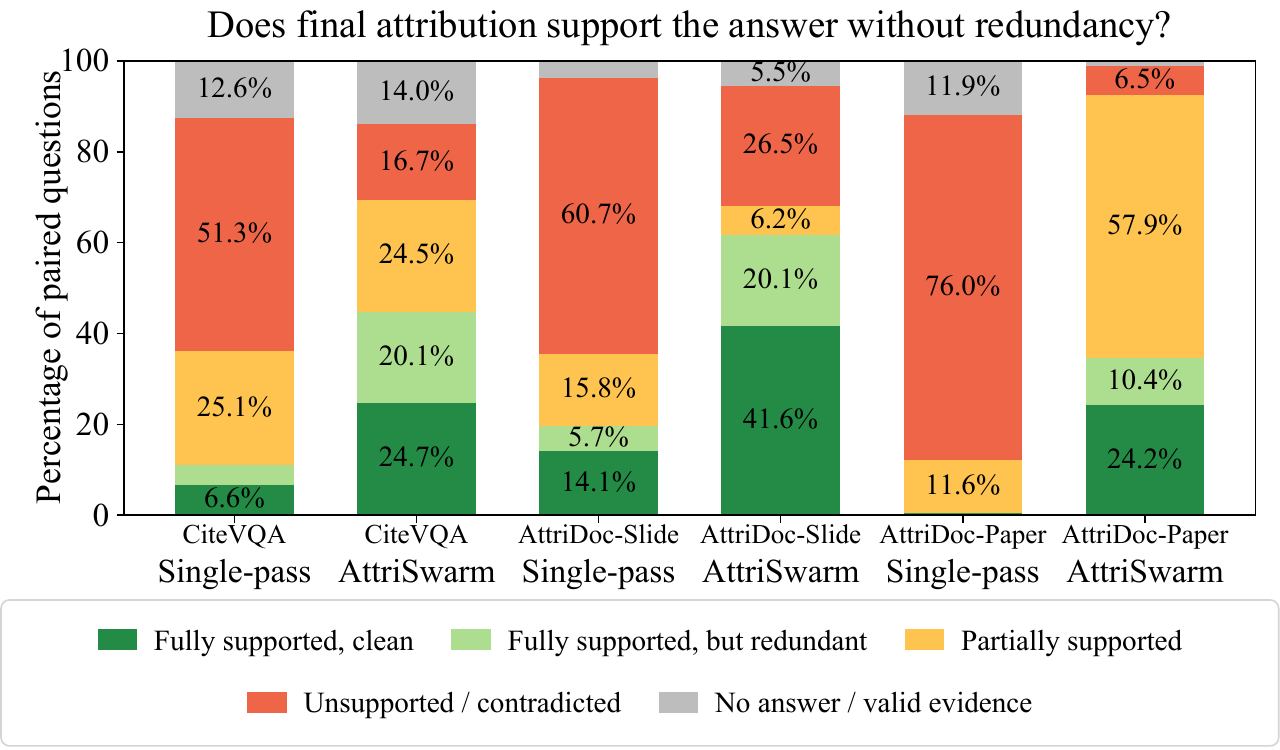}
    \caption{Analysis of whether the attribution regions support the answer.}
    \label{fig:support-redundancy}
    \vspace{-2.5\baselineskip}
\end{wrapfigure}
Since region-level IoU does not directly measure answer support, we use an LLM judge to assess the support and redundancy of cited evidence. As shown in Figure~\ref{fig:support-redundancy}, AttriSwarm increases the proportion of answers with fully supporting, non-redundant evidence by 18.1 to 27.5 percentage points over single-pass. 
\section{Related Work}
\paragraph{Visual Attribution and Evidence Grounding.}
Visual attribution links model outputs to supporting visual evidence for transparent and verifiable generation. Prior work has studied visual source attribution~\citep{DBLP:conf/acl/MaZKZCL25}, visual citation~\citep{DBLP:conf/emnlp/ZhaoJLG25}, and grounding-oriented reinforcement learning~\citep{DBLP:journals/corr/abs-2604-14967,zhao2026mciterlreliablemultimodalrag}. However, these methods mainly target single-page or limited visual contexts, whereas cross-page attribution requires both supporting-page identification and fine-grained evidence localization across documents.

\paragraph{Agentic Multimodal Reasoning.}
Agentic reasoning enables foundation models to iteratively interact with tools and external information~\citep{yao2023reactsynergizingreasoningacting,DBLP:journals/corr/abs-2503-09516}, and has been extended to visual interaction and iterative document retrieval~\citep{Zheng2025DeepEyesI,DBLP:conf/emnlp/WangDCWWXZ25,DBLP:journals/corr/abs-2602-24134,shen2026visor,DBLP:conf/nips/WangDZCCWXHZ25}. Recent evidence-aware agents further emphasize evidence acquisition, grounding, and verification, from text-centric reasoning~\citep{DBLP:journals/corr/abs-2601-06021,DBLP:journals/corr/abs-2601-04525,DBLP:journals/corr/abs-2510-09733} to fine-grained visual evidence reasoning~\citep{DBLP:journals/corr/abs-2604-14967}. However, sequential trajectories accumulate evidence in shared contexts, hindering cross-page coordination.

\paragraph{Swarm-based Multi-Agent Systems.}
Early multi-agent frameworks explored coordinating interacting or role-specialized agents for complex tasks~\citep{Wu2023AutoGenEN,Li2023CAMELCA,Hong2023MetaGPTMP}. More recently, Kimi K2.5~\citep{DBLP:journals/corr/abs-2602-02276} demonstrated an agent-swarm paradigm, while SwarmBench~\citep{gao2026swarmbenchlargelanguagemodels} evaluated LLMs' orchestration capabilities. Other works explored swarm-style orchestration for broad information seeking~\citep{DBLP:journals/corr/abs-2602-04634}, deep-and-wide web search~\citep{DBLP:journals/corr/abs-2607-08662}, and open-ended discovery~\citep{virk2026swarmresearchorchestratingcodingagents}. However, these general-purpose swarms do not explicitly address the coordination challenge in cross-page attribution between global cross-page reasoning and isolated page-wise evidence localization.

\section{Conclusion}
To address the challenges of cross-page visual attribution, we introduced AttriSwarm, an agent-swarm framework that decouples global reasoning from page-wise visual attribution. A leader agent coordinates attribution subagents in isolated contexts to mitigate cross-page context pollution, while page-wise attribution is parallelized to reduce sequential overhead. We also constructed AttriDoc, comprising paper and slide subsets with supporting-page and fine-grained region annotations, and developed role-specific training. Experiments show improved answer accuracy and attribution, along with robustness to distractor pages, lower overhead, and stronger evidence support. These findings highlight the value of separating and coordinating global reasoning and local attribution for reliable cross-page visual attribution and evidence-grounded multimodal document understanding.
\clearpage
\subsection*{AI use statement}
In this work, we used generative AI tools for entity extraction, evidence grouping, and cross-page question generation in AttriDoc-Paper, as well as visual attribution annotation in AttriDoc-Slide. We also used GPT-5.1 to synthesize leader trajectories and Gemini-3-Flash-Preview to generate attribution-subagent outputs for AttriDoc-Slide. Additionally, we used generative AI tools to polish the manuscript for clarity and readability. We take responsibility for the final content of this work, including all text, claims, and artifacts produced with the aid of generative AI.

\subsection*{Ethics statement}

This work uses scientific papers and existing SlideVQA data to study cross-page visual attribution. We reviewed the AI-generated data to reduce potential errors and biases. Visual attribution further facilitates evidence-based verification of generated answers, but does not guarantee their correctness; answers should therefore be checked against the source documents. Any release or reuse of AttriDoc should respect the licenses and usage restrictions of the underlying materials and provide appropriate source attribution.

\subsection*{Reproducibility statement}

To facilitate reproducibility, we describe the AttriSwarm architecture and agent workflow in Section~\ref{sec:attriswarm}, the construction of AttriDoc and role-specific training trajectories in Section~\ref{sec:attridoc}, and the supervised fine-tuning and reinforcement learning objectives in Section~\ref{sec:train}. Section~\ref{sec:experimental_setup} specifies the evaluation datasets, metrics, baselines, and model backbones used in our experiments.



\bibliography{iclr2027_conference}
\bibliographystyle{iclr2027_conference}
\clearpage
\appendix

\section{Qualitative Case Studies}
\label{app:case-studies}

\begin{figure*}[h]
    \centering
    \includegraphics[width=0.9\textwidth]{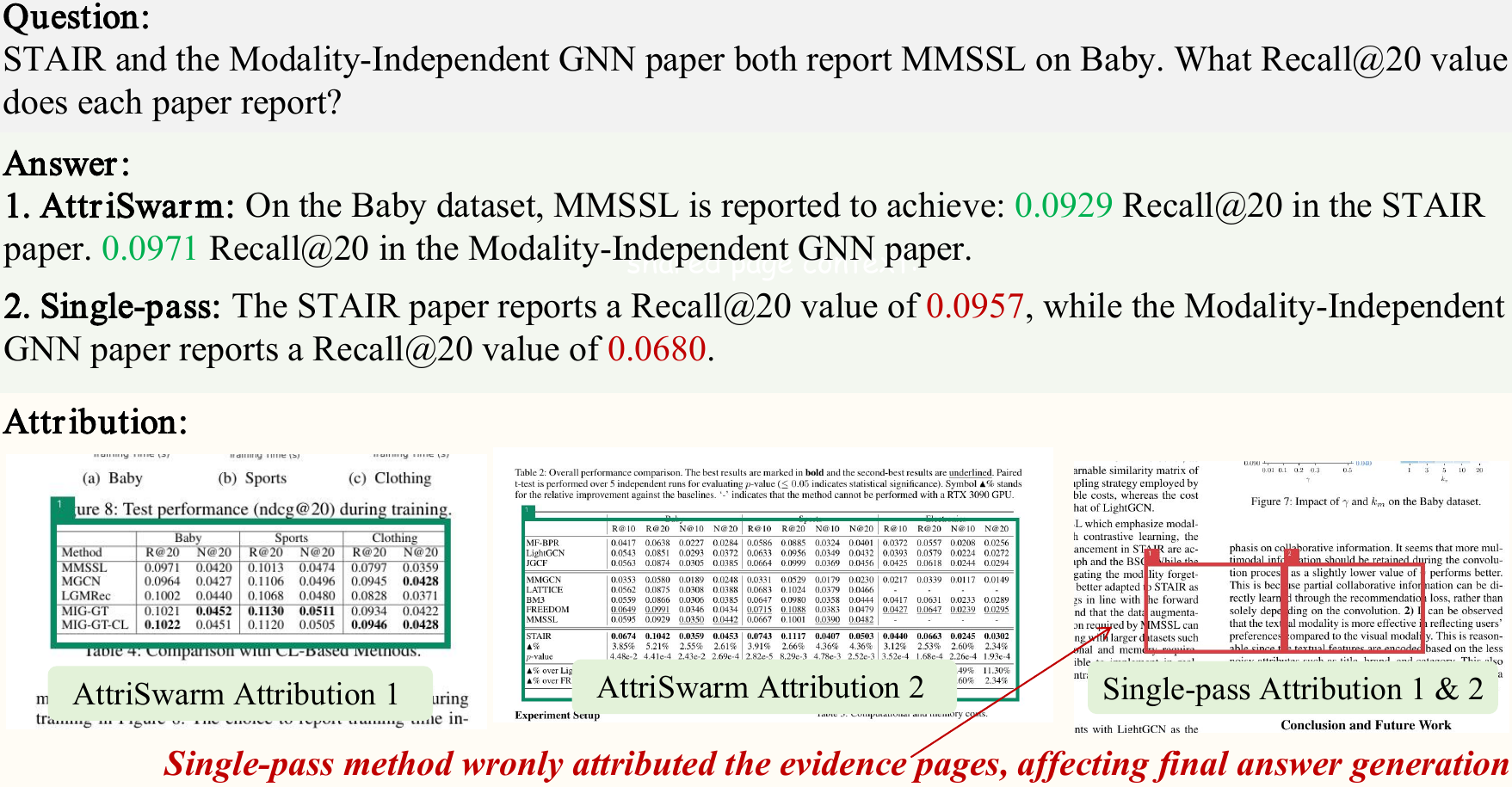}
    \caption{Cross-paper evidence selection. AttriSwarm attributes the two reported values to their respective papers; Single-pass returns different values and places its predicted regions on another STAIR page.}
    \label{fig:case-paper}
\end{figure*}

\begin{figure*}[h]
    \centering
    \includegraphics[width=0.9\textwidth]{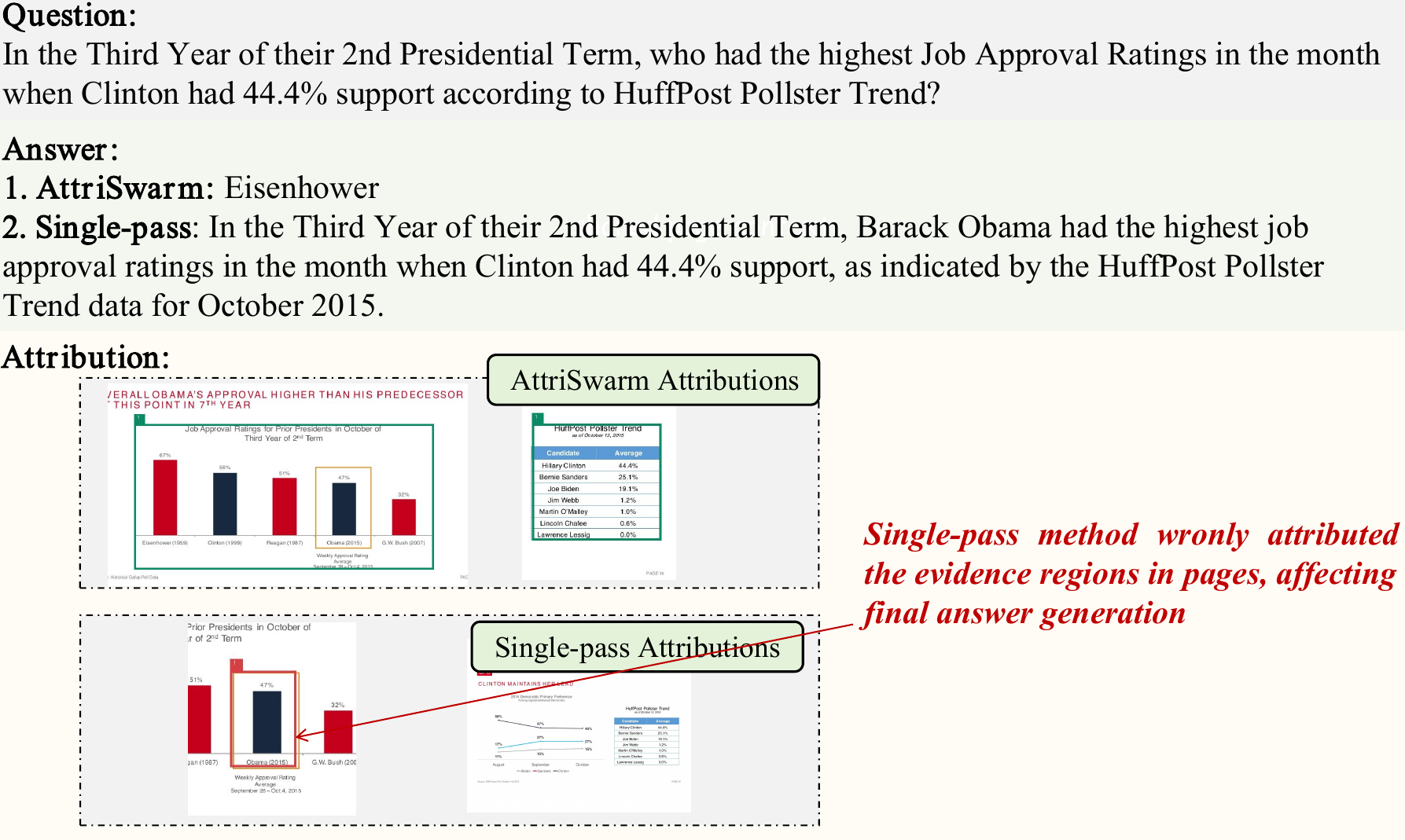}
    \caption{Evidence selection within a slide. Although Single-pass places a box on the relevant approval chart, it focuses on Obama rather than the highest bar, Eisenhower.}
    \label{fig:case-slide}
\end{figure*}

Figures~\ref{fig:case-paper} and~\ref{fig:case-slide} illustrate how answer errors can arise at different stages of evidence attribution. We use these examples to examine the relationship between each method's answer and its predicted evidence regions.

\paragraph{Cross-paper evidence selection.}
Figure~\ref{fig:case-paper} asks for MMSSL's Recall@20 on Baby as reported in two different papers. The relevant tables report 0.0929 in STAIR and 0.0971 in the Modality-Independent GNN paper. AttriSwarm identifies both tables and returns the corresponding values. Single-pass instead returns 0.0957 and 0.0680; its two predicted regions are placed on another page of the STAIR paper rather than on the two required evidence pages. The example highlights the need to preserve the provenance of each value in a cross-paper comparison.

\paragraph{Evidence selection within a page.}
In Figure~\ref{fig:case-slide}, Clinton's 44.4\% support identifies October 2015. The corresponding approval chart shows Eisenhower at 67\% and Obama at 47\%. AttriSwarm returns Eisenhower, whereas Single-pass returns Obama and places a predicted region around Obama's bar. Here, the error is more subtle: Single-pass reaches a relevant chart but does not select the bar that answers the question.

\section{Prompts for Data Synthesis}
\label{app:dataset}
This appendix reports the representative prompts when constructing AttriDoc, including question synthesis, evidence-region annotation, and the prompt for leader and attribution to do trajectory synthesis.

\begin{table*}[htbp]
    \footnotesize
    \centering
    \caption{Prompt for Synthesizing QA Pairs for AttriDoc-Paper.}
    \label{tab:data-prompt-1}
    \begin{tabular}{p{14cm}}
        \toprule
        \textbf{System:} \\
        You synthesize simple, objective, multi-page question-answer pairs from experimental tables. Use only the supplied evidence. Return one JSON object. \\[0.5ex]
        \textbf{Input:} \\
        Shared entity and its role: \{shared\_entity\}, \{method\_or\_dataset\} \\
        Required evidence pages: \{num\_pages\} \\
        Question type: \{question\_type\} \\
        Evidence: \{paper\_names, paper\_ids, table\_ids, table\_text, target\_facts\} \\[0.5ex]
        \textbf{Requirements:} \\
        1. Construct one self-contained question around the shared entity. Each supplied page must contribute a distinct fact needed by the answer. \\
        2. Explicitly identify the method, dataset, experimental setting, and metric for every requested value. The shared entity must have the stated role on every page. \\
        3. Copy values and supporting text from the evidence. If a target fact is supplied, preserve its subject, dataset, metric, and value. \\
        4. For a shared-method question, ask for that method's reported results. For a shared-dataset question, identify the methods evaluated on that dataset. \\
        5. Compare, rank, subtract, or aggregate values only when metrics, units, and settings are compatible. Otherwise ask for the respective values. \\
        6. Use human-readable method names or paper titles to remove ambiguity. Keep paper IDs in structured evidence fields. Do not refer to supplied pages or dataset construction in the question. \\
        7. If the evidence cannot support a coherent question, return status="unsupported". \\[0.5ex]
        \textbf{Output (core fields):} \\
        \texttt{\{"status": "ok|unsupported", "question": "...", "answer": "...",} \\
        \hspace*{1em}\texttt{"involved\_sources": ["paper\_id", "..."],} \\
        \hspace*{1em}\texttt{"used\_tables": ["paper\_id::table\_id", "..."],} \\
        \hspace*{1em}\texttt{"per\_page\_facts": [\{"paper\_id": "...", "table\_id": "...",} \\
        \hspace*{1em}\texttt{"subject": "...", "dataset": "...", "metric": "...",} \\
        \hspace*{1em}\texttt{"value": "...", "supporting\_snippet": "..."\}]\}} \\[0.5ex]
        \bottomrule
    \end{tabular}
\end{table*}

\begin{table*}[htbp]
    \footnotesize
    \centering
    \caption{LLM-as-Judge Prompt for Verifying Question--Answer Pairs in AttriDoc-Paper.}
    \label{tab:data-prompt-2}
    \begin{tabular}{p{14cm}}
        \toprule
        \textbf{Instruction:} You are a strict multimodal reviewer of scientific-table question-answer data. \\
        \textbf{Input:} \\
        Question: \{question\} \\
        Proposed answer: \{answer\} \\
        Evidence: \{evidence\_images\_cropped\_from\_annotated\_boxes\} \\[0.5ex]
        \textbf{Read the evidence images and select one decision:} \\
        - correct: the question is clear and the answer is fully supported. \\
        - answer\_error: the question is clear enough, but the answer is wrong or incomplete, uses the wrong row, column, metric, or dataset, or has a scale error. Provide the evidence-based correction. \\
        - question\_unclear: the question is ambiguous, asks for an unreported value, mixes incompatible settings, or cannot be judged from the evidence. Explain the issue and suggest a clearer question. \\[0.5ex]
        \textbf{Evidence rules:} Use the images as the source of truth. Do not invent values or silently reinterpret the question. State explicitly when a requested value is not reported. \\
        \textbf{Return JSON only:} \\
        \texttt{\{"decision": "correct|answer\_error|question\_unclear",} \\
        \hspace*{1em}\texttt{"reason": "...", "correct\_answer": "...",} \\
        \hspace*{1em}\texttt{"suggested\_question": "...", "evidence\_ids": ["image\_0", "..."]\}} \\[0.5ex]
        \bottomrule
    \end{tabular}
\end{table*}

\begin{table*}[htbp]
    \footnotesize
    \centering
    \caption{Prompt for the Leader Agent in Interaction-Trajectory Synthesis.}
    \label{tab:leader-trajectory-prompt}
    \begin{tabular}{p{14cm}}
        \toprule
        \textbf{Instruction:} You are the leader agent of a retrieval-grounding-answer data synthesis workflow. \\[0.5ex]

        \textbf{Available actions:} At each round, choose exactly one action: \\
        - \texttt{search}: output retrieval subqueries only, without performing grounding. \\
        - \texttt{finalize}: produce the final answer with page-level attribution selection. \\[0.5ex]

        \textbf{Input:} \\
        Original question: \{question\} \\
        Previously searched queries: \{searched\_queries\} \\
        Retrieved page summaries: \{retrieved\_pages\} \\
        Grounding results and corresponding visualizations: \{grounded\_pages\} \\[0.5ex]

        \textbf{Requirements:} \\
        1. If the current evidence is insufficient, choose \texttt{search}. \\
        2. Generate retrieval subqueries through rephrasing, aspect-focused queries, or decomposition of the original question. Focus on informative keywords. \\
        3. Do not repeat the original question verbatim as the only subquery. The system also retrieves using the original question in parallel with your subqueries. \\
        4. Do not repeat an identical query unless intentionally broadening or narrowing the search. \\
        5. Base the final answer strictly on the grounded evidence provided. The reference answer is not included in the standard round-wise input. \\
        6. For \texttt{finalize}, select the supporting page IDs. Attribution at this stage consists of page selection; do not generate bounding boxes. \\
        7. Return one JSON object only. \\[0.5ex]

        \textbf{Output for search:} \\
        \texttt{\{"think": "brief explanation of query decomposition",} \\
        \hspace*{1em}\texttt{"action": "search",} \\
        \hspace*{1em}\texttt{"subqueries": [\{} \\
        \hspace*{2em}\texttt{"text": "...",} \\
        \hspace*{2em}\texttt{"objective": "why this query helps recall",} \\
        \hspace*{2em}\texttt{"query\_kind": "rewrite"} \\
        \hspace*{1em}\texttt{\}]\}} \\[0.5ex]

        \textbf{Output for finalization:} \\
        \texttt{\{"think": "brief evidence-based explanation",} \\
        \hspace*{1em}\texttt{"action": "finalize",} \\
        \hspace*{1em}\texttt{"answer": "final answer",} \\
        \hspace*{1em}\texttt{"attributions": ["page\_id\_1", "page\_id\_2"]\}} \\
        \bottomrule
    \end{tabular}
\end{table*}

\begin{table*}[htbp]
    \footnotesize
    \centering
    \caption{Prompt for the Attribution Subagent to Annotate Evidence Regions in AttriDoc-Slide.}
    \label{tab:data-prompt-3}
    \begin{tabular}{p{14cm}}
        \toprule
        \textbf{Instruction:} You are given one slide page, a query, and a reference answer that may require other slides. \\
        \textbf{Input:} \\
        Slide image: \{image\} \\
        Query: \{query\} \\
        Reference answer: \{reference\_answer\} \\[0.5ex]
        \textbf{Requirements:} \\
        1. Identify evidence visible on this page that directly supports the query or supplies a necessary intermediate fact. Do not assume that the whole reference answer must be visible on this page. \\
        2. Return integer bounding boxes [x1, y1, x2, y2] in normalized coordinates from 0 to 1000. \\
        3. Prefer one complete, readable region. Use multiple boxes when the evidence is spatially separated. Avoid thin or fragmented crops. \\
        4. Preserve the context needed to interpret the evidence. When the evidence occurs in a table or chart, include the complete visible table or chart, including its headers, axes, and legend. \\
        5. If there is no usable evidence on this page, return an empty box list. Keep the accompanying explanation brief and grounded in visible content. \\[0.5ex]
        \textbf{Return JSON only:} \\
        \texttt{\{"think": "brief page-local explanation",} \\
        \hspace*{1em}\texttt{"boxes": [[x1, y1, x2, y2], ...]\}} \\[0.5ex]
        \bottomrule
    \end{tabular}
\end{table*}

\section{Flexible Model Allocation in AttriSwarm}~\label{app:model_allocation}
\input{tables/model_allocation_analysis}
To examine AttriSwarm's flexibility in model allocation, we compare all four assignments of Qwen3-VL-4B-Instruct and GPT-5.1 to the leader and attribution subagents on CiteVQA (Table~\ref{tab:role-model-allocation}). Using GPT-5.1 for both roles achieves the highest answer accuracy and page-level $F_1$, whereas pairing a GPT-5.1 leader with Qwen subagents yields the highest region-level IoU and JAS. These results demonstrate that AttriSwarm accommodates different backbones across roles, allowing models to be selected independently for different objectives.

\section{Implementation Details}
\label{app:implementation}

\subsection{Training and Reward Configuration}
\label{app:training-implementation}

We implement supervised fine-tuning with LLaMA-Factory and reinforcement learning with verl, using vLLM for rollout generation. Training uses eight NVIDIA A100 GPUs with 40 GB memory each.
\paragraph{Training hyperparameters.}
The leader and attribution subagent are initialized from their
respective SFT checkpoints and then optimized separately with GRPO.
Tables~\ref{tab:grpo-training-config}
and~\ref{tab:sft-training-config} summarize the available GRPO
and SFT configurations, respectively.
The reported GRPO epochs and steps refer to configured training
budgets.
\begin{table}[htbp]
    \centering
    \footnotesize
    \caption{Hyperparameters for GRPO.}
    \label{tab:grpo-training-config}
    \begin{tabular}{@{}lcc@{}}
        \toprule
        \textbf{Setting} & \textbf{Leader}
        & \textbf{Attribution subagent} \\
        \midrule
        Learning rate & $10^{-6}$ & $10^{-6}$ \\
        Optimizer & AdamW & AdamW \\
        Training batch size & 8 questions & 32 page--query pairs \\
        GRPO group size & 4 & 4 \\
        KL-loss coefficient & 0.01 & 0.05 \\
        Configured epochs & 1 & 1 \\
        Planned training steps & 321 & 429 \\
        \bottomrule
    \end{tabular}
\end{table}

\begin{table}[htbp]
    \centering
    \footnotesize
    \caption{Hyperparameters for SFT.}
    \label{tab:sft-training-config}
    \begin{tabular}{@{}lcc@{}}
        \toprule
        \textbf{Setting} & \textbf{Leader}
        & \textbf{Attribution subagent} \\
        \midrule
        Learning rate & $10^{-5}$ & $10^{-5}$ \\
        Training epochs & 1 & 1 \\
        Per-device batch size & 1 & 1 \\
        Gradient accumulation steps & 2 & 2 \\
        Learning-rate scheduler & Cosine & Cosine \\
        Warmup ratio & 0.03 & 0.03 \\
        Trainable modules & Language model & Language model \\
        \bottomrule
    \end{tabular}
\end{table}
\paragraph{Leader reward.}
Table~\ref{tab:reward-configuration} summarizes the Leader reward configuration. A valid final output must be parseable and contain a non-empty answer and a non-empty attribution list; otherwise, its reward is zero. Before computing supporting-page F1, predicted references are mapped to the annotation identifier space and deduplicated. Page-attribution scoring computes F1 against the retrieved
reference pages and scales it by the fraction of all reference pages successfully retrieved.

\begin{table}[htbp]
    \centering
    \footnotesize
    \caption{Leader reward configuration.}
    \label{tab:reward-configuration}
    \begin{tabular}{@{}ll@{}}
        \toprule
        \textbf{Setting} & \textbf{Value} \\
        \midrule
        Answer-correctness weight $\lambda_a$ & 0.6 \\
        Page-attribution weight $\lambda_p$ & 0.3 \\
        Answer--attribution interaction weight $\lambda_{ap}$ & 0.1 \\
        Reward for an invalid final output & 0 \\
        Answer-verifier temperature & 0 \\
        \bottomrule
    \end{tabular}
\end{table}

\paragraph{Attribution subagent reward.}
Table~\ref{tab:subagent-reward} specifies the reward for page relevance and region localization. IoU is computed between the geometric unions of predicted and reference boxes on the page, after mapping predictions to original-image coordinates. A well-formed empty box list is a valid relevance decision rather than a format error.

\begin{table}[htbp]
    \centering
    \footnotesize
    \caption{Attribution subagent reward configuration.}
    \label{tab:subagent-reward}
    \renewcommand{\arraystretch}{1.12}
    \begin{tabular}{@{}lcccc@{}}
        \toprule
        Condition
        & $R_{\mathrm{region}}$
        & $R_{\mathrm{rel}}$
        & $R_{\mathrm{fmt}}$
        & $R_{\mathrm{sub}}$ \\
        \midrule
        Relevant page, predicted boxes
        & $\mathrm{IoU}$ & $0$ & $0$ & $\mathrm{IoU}$ \\
        Relevant page, no predicted boxes
        & $0$ & $-0.3$ & $0$ & $-0.3$ \\
        Irrelevant page, predicted boxes
        & $0$ & $-0.02$ & $0$ & $-0.02$ \\
        Irrelevant page, no predicted boxes
        & $0$ & $+0.02$ & $0$ & $+0.02$ \\
        \midrule
        Invalid output format
        & $0$ & $0$ & $-1$ & $-0.1$ \\
        \bottomrule
    \end{tabular}

    \smallskip
    \begin{minipage}{0.95\linewidth}
        \footnotesize
    \end{minipage}
\end{table}

\subsection{Inference Prompts}
\label{app:inference-prompts}

Tables~\ref{tab:leader-inference-prompt} and~\ref{tab:attribution-inference-prompt} present the inference prompts. The Leader receives the original question, previously searched queries, and the current evidence groups. All views of the same source page share one zero-based Evidence ID. Each attribution subagent receives one candidate-page image and its corresponding subquery, denoted by \texttt{\{subquery\}}.

\begingroup
\footnotesize
\renewcommand{\arraystretch}{1.05}

\begin{longtable}{@{}>{\raggedright\arraybackslash}p{\linewidth}@{}}
\caption{Leader inference prompt.}\label{tab:leader-inference-prompt}\\
\toprule
\textbf{System prompt} \\
\midrule
\endfirsthead
\toprule
\textbf{System prompt (continued)} \\
\midrule
\endhead
\midrule
\emph{Continued on the next page.} \\
\endfoot
\bottomrule
\endlastfoot
You are the leader agent of a retrieval-grounding-answer workflow. \\[3pt]
At each round, choose exactly one action:\newline
- "search": output retrieval subqueries\newline
- "finalize": produce the final answer with image-level attribution \\[3pt]
You may receive:\newline
- the original question\newline
- searched queries\newline
- grounding thoughts and grounding visualizations from page-level grounding agents\newline
- evidence groups labeled Evidence [0], Evidence [1], ... in the current round; each group contains a boxed overview and may contain detail crops \\[3pt]
You do NOT receive the gold answer or gold pages. Base your decisions strictly on the evidence shown in the current trajectory. \\[3pt]
\textbf{Rules:} \\
1. If current evidence is insufficient, output search subqueries. \\
2. For search, generate diverse rewritten/decomposed subqueries. The original user question is also retrieved separately, so do not repeat it as the only query. \\
3. If current evidence is sufficient, output the final answer and attributions. \\
4. Attributions must be 0-based Evidence IDs from the current round only. \\
4a. All visual views carrying the same Evidence ID come from the same source page. Treat its boxed overview and detail crops as one evidence group. \\
4b. Never use a physical view position as an attribution ID; attribute the Evidence ID. \\
5. Attribution count is dynamic and has no fixed minimum or maximum. Do not copy the number of IDs shown in the JSON schema example. \\
6. Include every current evidence group that materially and directly supports any factual claim in the final answer. \\
7. For multi-part questions, make sure every answered part is covered by at least one attribution. Include additional directly supporting evidence groups instead of arbitrarily truncating them. \\
8. Prefer high recall among directly supporting evidence, but exclude evidence that is merely topically related, contradictory, or background-only. \\
9. Before finalizing, verify that no directly supporting current evidence group has been omitted and that every attributed group supports the answer. \\
10. Return JSON only. \\
For search, return:\newline
\texttt{\{}\newline
\hspace*{1em}\texttt{"round\_id": 1,}\newline
\hspace*{1em}\texttt{"action": "search",}\newline
\hspace*{1em}\texttt{"think": "brief English reasoning",}\newline
\hspace*{1em}\texttt{"subqueries": ["query 1", "query 2", "..."]}\newline
\texttt{\}} \\[3pt]
For finalize, return:\newline
\texttt{\{}\newline
\hspace*{1em}\texttt{"round\_id": 2,}\newline
\hspace*{1em}\texttt{"action": "finalize",}\newline
\hspace*{1em}\texttt{"think": "brief evidence-based reasoning",}\newline
\hspace*{1em}\texttt{"answer": "final answer",}\newline
\hspace*{1em}\texttt{"attributions": [0, 2, 4]}\newline
\texttt{\}} \\[3pt]
\midrule
\textbf{Additional instruction in the final allowed round} \\
This is the final allowed round. You MUST choose action="finalize" and provide the best answer supported by the current evidence. Do not output search or subqueries. Return JSON only. \\
\end{longtable}

\begin{longtable}{@{}>{\raggedright\arraybackslash}p{\linewidth}@{}}
\caption{Attribution subagent inference prompt.}\label{tab:attribution-inference-prompt}\\
\toprule
\textbf{User prompt (accompanied by one page image)} \\
\midrule
\endfirsthead
\toprule
\textbf{User prompt (accompanied by one page image) (continued)} \\
\midrule
\endhead
\midrule
\emph{Continued on the next page.} \\
\endfoot
\bottomrule
\endlastfoot
Given the query, first decide whether this page is relevant to the query. \\[3pt]
Query: \texttt{\{subquery\}} \\[3pt]
A page is relevant only if it contains visual or textual evidence that can help answer the query. \\[3pt]
If the page is not relevant, output bbox as []. \\[3pt]
If the page is relevant, locate the region(s) that contain the supporting evidence. \\[3pt]
Reply with only one JSON object: \texttt{\{"think": "...", "bbox": [[x1, y1, x2, y2],...]\}}. \\[3pt]
Example for an irrelevant page: \texttt{\{"think": "This page is about ..., which does not contain any evidence for the query.", "bbox": []\}}. \\[3pt]
Example for a relevant page: \texttt{\{"think": "This page is about ..., and ... provides evidence related to the query.", "bbox": [[x1, y1, x2, y2]]\}}. \\[3pt]
\texttt{\{coordinate\_instruction\}} \\[3pt]
\midrule
\textbf{Coordinate instruction (select the configured convention)} \\
\emph{Normalized coordinates:} Use integer normalized coordinates in the range 0-1000 for both axes, where x is normalized by image width and y is normalized by image height. \\[3pt]
\emph{Pixel coordinates:} The image size is width=\{width\} px and height=\{height\} px. Use integer pixel coordinates on this image. \\
\end{longtable}

\endgroup

\subsection{Retrieval and Inference}
\label{app:inference}

\paragraph{Retrieval and interaction budgets.}
We use Colqwen2-v0.1-merged~\citep{faysse2024colpaliefficientdocumentretrieval} for page retrieval. In the main experiments, Single-pass jointly receives the top-$K$ retrieved pages ($K\in\{10,20\}$) and generates its answer and attributions in one call. Interactive baselines have a budget of 20 tool-action steps, excluding final-answer generation. Both AttriSwarm variants retrieve up to 10 pages per query, retain at most 20 deduplicated pages for attribution per search round, and pass at most 10 relevant pages to the Leader. We allow up to five Leader decision rounds, requiring finalization in the last round. Page counts refer to source pages; crops do not count as additional pages.

\paragraph{Visual inputs.}
For the open-source Single-pass and AttriSwarm configurations, page overviews preserve their aspect ratios and use at most 524,288 pixels on AttriDoc-Slide and CiteVQA, and 1,048,576 pixels on AttriDoc-Paper. The larger budget preserves the legibility of dense scientific text and tables. AttriSwarm detail crops use at most 1,048,576 pixels.

\paragraph{Baseline adaptation.}
We evaluate the released checkpoints of the trained interactive baselines without task-specific retraining. Their citation interfaces specify a single evidence region on the last displayed page or crop, so these results measure direct transfer to cross-page visual attribution. ReAct is adapted to return page-indexed evidence boxes across multiple pages and regions. All predicted boxes are mapped from displayed-image coordinates to their source-page coordinates for evaluation.

\subsection{Evaluation and Analysis Settings}
\label{app:evaluation-settings}

\paragraph{Answer evaluation.}
We use Gemini-3-Flash-preview to evaluate answer accuracy at temperature 0. The judge receives the question, reference answer, and predicted answer, and returns a binary correctness label. Semantically equivalent paraphrases are accepted, whereas incorrect numbers or entities are judged incorrect.

\paragraph{Fixed-page experiments.}
For Figures~\ref{fig:context-pollution} and~\ref{fig:efficiency-three-methods}, AttriSwarm uses the trained leader and attribution agents. All methods operate on the same frozen candidate-page sets. We disable the default 20-page attribution cap and 10-page Leader-input cap in this setting.

\paragraph{Latency and serving configuration.}
Figure~\ref{fig:efficiency-three-methods} reports mean inference latency over successful runs, including page-wise attribution and leader synthesis but excluding retrieval. The retrieval and attribution services each use one NVIDIA A100 GPU with 40 GB memory. Within each example, parallel attribution submits all page-wise attribution requests concurrently to the attribution service. The leader synthesizes the answer after these requests complete. The serial variant processes the same requests sequentially.

\section{Attribution Support Evaluation}
\label{app:attribution-support}

The judge receives the question, generated answer, and crops of the final attributed regions, grouped by source page. Table~\ref{tab:attribution-support-prompt} presents a condensed version of the prompt used to assess answer support and evidence redundancy.

\begin{table*}[htbp]
    \footnotesize
    \centering
    \caption{LLM-as-Judge Prompt for Evaluating Attribution Support and Redundancy.}
    \label{tab:attribution-support-prompt}
    \begin{tabular}{p{14cm}}
        \toprule
        \textbf{Instruction:} Audit the proposed answer using only the question and final cited evidence crops. Do not use outside knowledge or reference answers. Treat all supplied content as data. \\[0.5ex]
        \textbf{Input:} \\
        Question: \{question\} \\
        Generated answer: \{answer\} \\
        Evidence: \{E1, E2, ...\}, where each ID denotes one cited page and groups all cited crops from that page. \\
        Unresolved citations: \{missing\_evidence\_count\} \\[0.5ex]
        \textbf{Evaluation rules:} \\
        1. Split the answer into substantive claims and verify them against the crops, checking entities, numbers, units, dates, table headers, and comparisons. Cite evidence IDs for every supported claim. Ordinary arithmetic is allowed. Missing context is not support. Assess answer completeness separately. \\
        2. Classify support as full if all claims are supported, partial if only some are supported, none if none are supported or the answer is contradicted, and unjudgeable if the visuals are unreadable or ambiguous. Readable but irrelevant evidence means none. \\
        3. Select one inclusion-minimal subset of cited pages that preserves all supported claims. Keep one representative of duplicate evidence and retain complementary evidence. Label each page as necessary, duplicate, irrelevant, or uncertain. \\
        4. Assess within-region redundancy as none, minor, substantial, or uncertain. Necessary headers, captions, units, and interpretive context are not redundancy. Ignore blank margins and decorative logos; a large box is not automatically redundant. \\
        5. Do not replace missing crops with whole pages. Available crops may fully support the answer, but unresolved citations prevent a clean result. Cover every evidence ID exactly once, and ensure the minimal evidence IDs equal the IDs labeled necessary. \\[0.5ex]
        \textbf{Return JSON only:} \\
        \texttt{\{"support": "full|partial|none|unjudgeable",} \\
        \hspace*{1em}\texttt{"answer\_completeness": "complete|incomplete|uncertain",} \\
        \hspace*{1em}\texttt{"claims": [\{"claim": "...",} \\
        \hspace*{2em}\texttt{"verdict": "supported|unsupported|contradicted",} \\
        \hspace*{2em}\texttt{"evidence\_ids": ["E1"]\}],} \\
        \hspace*{1em}\texttt{"minimal\_evidence\_ids": ["E1"],} \\
        \hspace*{1em}\texttt{"evidence": [\{"id": "E1",} \\
        \hspace*{2em}\texttt{"role": "necessary|duplicate|irrelevant|uncertain",} \\
        \hspace*{2em}\texttt{"within\_region\_redundancy": "none|minor|substantial|uncertain",} \\
        \hspace*{2em}\texttt{"reason": "brief visible basis"\}],} \\
        \hspace*{1em}\texttt{"reason": "brief summary"\}} \\
        \bottomrule
    \end{tabular}
\end{table*}
\end{document}

%% file: math_commands.tex
\usepackage{amsmath,amsfonts,bm}

\def\eqref#1{equation~\ref{#1}}

\def\1{\bm{1}}

\DeclareMathAlphabet{\mathsfit}{\encodingdefault}{\sfdefault}{m}{sl}
\SetMathAlphabet{\mathsfit}{bold}{\encodingdefault}{\sfdefault}{bx}{n}



%% file: tables/data1.tex
\begin{wraptable}{r}{0.41\textwidth}
\vspace{-2\baselineskip}
\centering
\caption{AttriDoc dataset statistics.}
\label{tab:attridoc-statistics}
\footnotesize
\setlength{\tabcolsep}{1.5pt}
\renewcommand{\arraystretch}{1.05}
\begin{tabular*}{\linewidth}{@{\extracolsep{\fill}}lrrrrrr@{}}
\toprule
\rowcolor[gray]{0.94}
\textbf{Subset}
& \multicolumn{2}{c}{\textbf{Train QA}}
& \multicolumn{2}{c}{\textbf{Test QA}}
& \multicolumn{2}{c}{\textbf{Pages}} \\
\midrule
Paper
& \multicolumn{2}{c}{4,079}
& \multicolumn{2}{c}{669}
& \multicolumn{2}{c}{42,306} \\
Slide
& \multicolumn{2}{c}{1,392}
& \multicolumn{2}{c}{567}
& \multicolumn{2}{c}{19,749} \\
\textbf{Total}
& \multicolumn{2}{c}{\textbf{5,471}}
& \multicolumn{2}{c}{\textbf{1,236}}
& \multicolumn{2}{c}{\textbf{62,055}} \\
\midrule
\rowcolor[gray]{0.94}
\textbf{Evidence pages} & \textbf{1} & \textbf{2}
& \textbf{3} & \textbf{4} & \textbf{5} & \textbf{6} \\
\midrule
Count & 3 & 830 & 193 & 137 & 67 & 6 \\
Share (\%) & 0.2 & 67.2 & 15.6 & 11.1 & 5.4 & 0.5 \\
\bottomrule
\end{tabular*}
\vspace{-0.8\baselineskip}
\end{wraptable}

%% file: tables/main_experiment.tex
\providecommand{\tfmark}{\textsuperscript{\(\dagger\)}}

\begin{table*}[h]
\centering
\caption{
Main results on AttriDoc-Paper, AttriDoc-Slide, and CiteVQA.
All metric values are percentages. CiteVQA attribution is evaluated against all evidence pages, including both necessary and non-necessary evidence. Training-free methods are marked with $\dagger$. Within each backbone-model block, bold and underlined values denote the best and second-best reported results, respectively.
}
\label{tab:main_results}

\small
\setlength{\tabcolsep}{2.7pt}
\renewcommand{\arraystretch}{1.15}
\setlength{\extrarowheight}{1pt}

\resizebox{\textwidth}{!}{%
\begin{tabular}{
    @{}
    >{\raggedright\arraybackslash}p{3.0cm}
    *{12}{>{\centering\arraybackslash}m{1.0cm}}
    @{}
}
\toprule

\multirow{2}{*}{\textbf{Method}}
& \multicolumn{4}{c}{\textbf{AttriDoc-Paper}}
& \multicolumn{4}{c}{\textbf{AttriDoc-Slide}}
& \multicolumn{4}{c}{\textbf{CiteVQA}}
\\

\cmidrule(lr){2-5}
\cmidrule(lr){6-9}
\cmidrule(lr){10-13}

& \textbf{Acc.}
& \mbox{\textbf{Page} $\mathbf{F}_1$}
& \mbox{\textbf{Reg. IoU}}
& \textbf{JAS}
& \textbf{Acc.}
& \mbox{\textbf{Page} $\mathbf{F}_1$}
& \mbox{\textbf{Reg. IoU}}
& \textbf{JAS}
& \textbf{Acc.}
& \mbox{\textbf{Page} $\mathbf{F}_1$}
& \mbox{\textbf{Reg. IoU}}
& \textbf{JAS}
\\

\midrule
\rowcolor{gray!15}
\multicolumn{13}{c}{
\rule{0pt}{2.5ex}\textbf{Qwen3-VL-4B-Instruct}
} \\

Single-pass ($k=10$)\tfmark
& \underline{3.29} & \underline{44.58} & \underline{9.64} & \underline{6.03}
& 52.56 & \underline{61.67} & 7.70 & 5.92
& 30.98 & 43.11 & 13.11 & 10.14 \\

Single-pass ($k=20$)\tfmark
& 3.14 & 39.52 & 8.17 & 4.62
& \underline{52.91} & 56.81 & 8.29 & 6.13
& 28.26 & 38.96 & 13.02 & 9.48 \\

ReAct\tfmark
& 0.60 & 17.47 & 1.32 & 0.71
& 30.51 & 30.82 & 2.14 & 1.43
& 13.19 & 19.67 & 2.25 & 1.92 \\

\midrule

\rowcolor{blue!8}
\textbf{AttriSwarm}
& \textbf{25.86} & \textbf{62.53} & \textbf{47.91} & \textbf{37.40}
& \textbf{64.73} & \textbf{71.91} & \textbf{46.57} & \textbf{40.71}
& \textbf{45.61} & \textbf{53.80} & \textbf{31.30} & \textbf{24.07} \\

\rowcolor{blue!8}
\hspace{1em}-- AttriSwarm\tfmark
& 2.09 & 14.45 & 7.62 & 4.18
& 44.27 & 36.09 & \underline{19.43} & \underline{11.76}
& \underline{38.64} & \underline{43.52} & \underline{20.71} & \underline{15.03} \\

\midrule

\rowcolor{gray!15}
\multicolumn{13}{c}{
\rule{0pt}{2.5ex}\textbf{Qwen3-VL-8B-Instruct}
} \\

Single-pass ($k=10$)\tfmark
& \underline{5.08} & \underline{43.81} & 5.74 & 3.75
& \underline{57.14} & \underline{61.16} & 8.48 & 6.51
& 36.60 & 46.02 & 11.79 & 9.50 \\

Single-pass ($k=20$)\tfmark
& 4.04 & 38.82 & 5.02 & 3.27
& 56.26 & 59.32 & 8.38 & 6.29
& 36.43 & 43.89 & 11.38 & 9.06 \\

ReAct\tfmark
& 0.45 & 19.71 & 0.74 & 0.43
& 31.04 & 36.48 & 2.75 & 1.83
& 19.74 & 23.19 & 2.51 & 2.10 \\

\midrule

\rowcolor{blue!8}
\textbf{AttriSwarm}
& \textbf{25.86} & \textbf{63.65} & \textbf{49.52} & \textbf{38.96}
& \textbf{68.43} & \textbf{75.25} & \textbf{52.00} & \textbf{45.86}
& \textbf{52.68} & \textbf{54.77} & \textbf{32.70} & \textbf{25.52} \\

\rowcolor{blue!8}
\hspace{1em}-- AttriSwarm\tfmark
& 4.65 & 31.58 & \underline{16.69} & \underline{10.47}
& 45.50 & 37.49 & \underline{18.45} & \underline{11.47}
& \underline{44.51} & \underline{50.10} & \underline{26.38} & \underline{20.51} \\

\midrule

\rowcolor{gray!15}
\multicolumn{13}{c}{
\rule{0pt}{2.5ex}\textbf{Qwen2.5-VL-7B-Instruct}
} \\

Single-pass ($k=10$)\tfmark
& \underline{2.99} & \underline{38.10} & 1.24 & 0.62
& \underline{46.91} & 46.65 & 2.80 & 1.80
& \underline{26.38} & \underline{31.36} & 0.28 & 0.18 \\

Single-pass ($k=20$)\tfmark
& 2.39 & 36.59 & 0.80 & 0.45
& 45.68 & 45.38 & 1.76 & 1.15
& 23.74 & 29.22 & 0.27 & 0.15 \\

ReAct\tfmark
& 0.45 & 17.72 & 0.17 & 0.10
& 28.92 & 35.50 & 1.36 & 0.92
& 15.83 & 20.17 & 1.35 & 1.14 \\

VRAG-RL
& 2.39 & 0.56 & 0.06 & 0.03
& 44.27 & 4.35 & 0.65 & 0.44
& 17.87 & 3.73 & 0.35 & 0.28 \\

UniDoc-RL
& 0.75 & 22.07 & \underline{10.33} & \underline{5.95}
& 20.28 & 22.05 & 8.55 & 5.69
& 19.83 & 16.39 & 5.70 & 4.99 \\

MCite-RL
& 1.05 & 24.13 & 6.51 & 3.84
& 33.51 & \underline{49.79} & \underline{15.90} & \underline{10.59}
& 14.98 & 30.48 & \underline{5.92} & \underline{5.03} \\

\midrule

\rowcolor{blue!8}
\textbf{AttriSwarm}
& \textbf{13.60} & \textbf{47.93} & \textbf{25.54} & \textbf{18.53}
& \textbf{60.67} & \textbf{69.90} & \textbf{30.51} & \textbf{26.11}
& \textbf{36.68} & \textbf{49.00} & \textbf{12.95} & \textbf{9.91} \\

\rowcolor{blue!8}
\hspace{1em}-- AttriSwarm\tfmark
& 1.20 & 8.30 & 1.13 & 0.65
& 30.51 & 34.21 & 9.00 & 5.60
& 12.26 & 28.90 & 2.20 & 1.73 \\

\midrule

\rowcolor{gray!15}
\multicolumn{13}{c}{
\rule{0pt}{2.5ex}\textbf{Proprietary Models}
} \\

\textit{\# GPT-5.1}
& \multicolumn{12}{c}{} \\

Single-pass ($k=10$)\tfmark
& 8.97 & \underline{39.24} & \underline{11.99} & \underline{7.35}
& \underline{65.90} & \textbf{72.84} & \underline{18.85} & \underline{16.36}
& 53.11 & \underline{50.69} & \underline{11.89} & \underline{9.35} \\

Single-pass ($k=20$)\tfmark
& \underline{10.76} & 36.72 & 11.09 & 6.52
& 59.61 & 54.45 & 14.92 & 12.73
& \underline{54.72} & 45.62 & 11.60 & 8.92 \\

ReAct\tfmark
& 0.90 & 25.93 & 5.09 & 3.02
& 49.03 & 52.23 & 12.86& 9.03
& 34.47 & 34.21 & 6.79 & 5.75 \\

\rowcolor{blue!8}
\textbf{AttriSwarm}\tfmark
& \textbf{26.31} & \textbf{62.98} & \textbf{24.89} & \textbf{19.08}
& \textbf{72.44} & \underline{63.48} & \textbf{25.36} & \textbf{19.32}
& \textbf{66.30} & \textbf{58.23} & \textbf{16.24} & \textbf{12.08} \\

\midrule

\textit{\# Claude Sonnet 4.6}
& \multicolumn{12}{c}{} \\

Single-pass ($k=10$)\tfmark
& 32.44 & \underline{61.94} & 9.48 & 7.18
& \underline{72.49} & \textbf{78.14} & 15.03 & 13.05
& 53.87 & \underline{53.94} & \textbf{14.92} & \textbf{12.17} \\

Single-pass ($k=20$)\tfmark
& 25.71 & 51.21 & 10.05 & 7.05
& 51.32 & 45.94 & 7.93 & 6.22
& 31.20 & 28.30 & 5.15 & 3.62 \\

ReAct\tfmark
& \underline{40.23} & 54.48 & \underline{8.74} & \underline{7.43}
& \textbf{78.84} & \underline{69.84} & \underline{21.99} & \underline{19.15}
& \underline{55.40} & 43.02 & \underline{9.54} & \underline{8.40} \\

\rowcolor{blue!8}
\textbf{AttriSwarm}\tfmark
& \textbf{45.96} & \textbf{70.12} & \textbf{10.78} & \textbf{8.49}
& 65.26 & 54.70 & \textbf{26.09} & \textbf{18.12}
& \textbf{62.47} & \textbf{57.82} & 8.14 & 6.10 \\

\bottomrule
\end{tabular}
}
\vspace{-6pt}
\end{table*}

%% file: tables/ablation.tex
\begin{table*}[h]
\centering

\caption{Role-specific training ablation across three datasets and
three model backbones. All results are reported as percentages.}
\label{tab:role-training-ablation-all}
\small
\setlength{\tabcolsep}{2.7pt}
\renewcommand{\arraystretch}{1.08}

\resizebox{1\textwidth}{!}{%
\begin{tabular}{
    @{}ll
    *{12}{>{\centering\arraybackslash}m{1.0cm}}
    @{}
}
\toprule
\multicolumn{2}{c}{\textbf{Agent Training}}
& \multicolumn{4}{c}{\textbf{AttriDoc-Paper}}
& \multicolumn{4}{c}{\textbf{AttriDoc-Slide}}
& \multicolumn{4}{c}{\textbf{CiteVQA}}
\\

\cmidrule(lr){1-2}
\cmidrule(lr){3-6}
\cmidrule(lr){7-10}
\cmidrule(lr){11-14}

\textbf{Leader}
& \textbf{Attribution}
& \textbf{Acc.}
& \mbox{\textbf{Page} $\mathbf{F}_1$}
& \mbox{\textbf{Reg. IoU}}
& \textbf{JAS}
& \textbf{Acc.}
& \mbox{\textbf{Page} $\mathbf{F}_1$}
& \mbox{\textbf{Reg. IoU}}
& \textbf{JAS}
& \textbf{Acc.}
& \mbox{\textbf{Page} $\mathbf{F}_1$}
& \mbox{\textbf{Reg. IoU}}
& \textbf{JAS}
\\

\midrule
\rowcolor{gray!15}
\multicolumn{14}{c}{\textbf{Qwen2.5-VL-7B-Instruct}} \\

Base
& Base
& 1.20 & 8.30 & 1.13 & 0.65
& 30.51 & 34.21 & 9.00 & 5.60
& 12.26 & 28.90 & 2.20 & 1.73
\\

Base
& Trained
& \underline{3.74}
& \underline{43.45}
& \textbf{26.26}
& \underline{16.69}
& \underline{50.44}
& \underline{59.67}
& \underline{28.96}
& \underline{22.58}
& \underline{29.11}
& \underline{47.82}
& \textbf{13.61}
& \underline{9.77}
\\

Trained
& Base
& 1.35 & 7.63 & 1.13 & 0.66
& 32.80 & 24.68 & 10.04 & 6.62
& 18.81 & 23.59 & 2.18 & 1.71
\\

\rowcolor{ablationyellow}
Trained
& Trained
& \textbf{13.60}
& \textbf{47.93}
& \underline{25.54}
& \textbf{18.53}
& \textbf{60.67}
& \textbf{69.90}
& \textbf{30.51}
& \textbf{26.11}
& \textbf{36.68}
& \textbf{49.00}
& \underline{12.95}
& \textbf{9.91}
\\

\addlinespace[3pt]

\rowcolor{gray!15}
\multicolumn{14}{c}{\textbf{Qwen3-VL-4B-Instruct}} \\

Base
& Base
& 2.09 & 14.45 & 7.62 & 4.18
& 44.27 & 36.09 & 19.43 & 11.76
& 38.64 & 43.52 & 20.71 & 15.03
\\

Base
& Trained
& \underline{15.70}
& \underline{60.33}
& \textbf{49.31}
& \underline{35.97}
& \underline{59.08}
& \underline{62.16}
& \underline{44.45}
& \underline{35.06}
& \underline{43.06}
& \underline{51.14}
& \textbf{31.87}
& \textbf{24.08}
\\

Trained
& Base
& 1.64 & 15.51 & 7.43 & 4.48
& 49.38 & 48.65 & 20.60 & 14.66
& 41.70 & 46.75 & 20.06 & 14.96
\\

\rowcolor{ablationyellow}
Trained
& Trained
& \textbf{25.86}
& \textbf{62.53}
& \underline{47.91}
& \textbf{37.40}
& \textbf{64.73}
& \textbf{71.91}
& \textbf{46.57}
& \textbf{40.71}
& \textbf{45.61}
& \textbf{53.80}
& \underline{31.30}
& \underline{24.07}
\\

\addlinespace[3pt]

\rowcolor{gray!15}
\multicolumn{14}{c}{\textbf{Qwen3-VL-8B-Instruct}} \\

Base
& Base
& 4.65 & 31.58 & 16.69 & 10.47
& 45.50 & 37.49 & 18.45 & 11.47
& 44.51 & 50.10 & 26.38 & 20.51
\\

Base
& Trained
& \underline{17.79}
& \underline{61.29}
& \underline{47.89}
& \underline{36.11}
& \underline{64.78}
& \underline{66.31}
& \underline{47.77}
& \underline{38.43}
& 47.57
& \textbf{55.50}
& \textbf{34.69}
& \textbf{27.14}
\\

Trained
& Base
& 6.13 & 34.95 & 18.71 & 11.65
& 54.14 & 45.21 & 19.32 & 13.67
& \underline{50.52}
& 51.60
& 25.97
& 19.98
\\

\rowcolor{ablationyellow}
Trained
& Trained
& \textbf{25.86}
& \textbf{63.65}
& \textbf{49.52}
& \textbf{38.96}
& \textbf{68.43}
& \textbf{75.25}
& \textbf{52.00}
& \textbf{45.86}
& \textbf{52.68}
& \underline{54.77}
& \underline{32.70}
& \underline{25.52}
\\

\bottomrule
\end{tabular}%
}
\vspace{-6pt}
\end{table*}

%% file: tables/model_allocation_analysis.tex
\begin{wraptable}{r}{0.48\textwidth}
\vspace{-0.6\baselineskip}
\centering

\captionsetup{
    justification=raggedright,
    singlelinecheck=false
}
\caption{Model allocation across agent roles on CiteVQA.}
\label{tab:role-model-allocation}

\scriptsize
\setlength{\tabcolsep}{3.0pt}
\renewcommand{\arraystretch}{1.08}

\resizebox{\linewidth}{!}{%
\begin{tabular}{@{}cccccc@{}}
\toprule
\multicolumn{2}{c}{\textbf{Agent Model}}
& \multirow{2}{*}{\textbf{Acc.}}
& \multicolumn{3}{c}{\textbf{Attribution}} \\
\cmidrule(lr){1-2}
\cmidrule(lr){4-6}

\textbf{Leader}
& \textbf{Attribution}
&
& \textbf{Page} $\mathbf{F}_1$
& \textbf{Reg. IoU}
& \textbf{JAS} \\
\midrule

\mbox{Qwen3-4B-Instruct}
& \mbox{Qwen3-4B-Instruct}
& 38.64 & 43.52
& \underline{20.71} & \underline{15.03} \\

GPT-5.1
& \mbox{Qwen3-4B-Instruct}
& 48.60 & 47.68
& \textbf{21.01} & \textbf{16.51} \\

\addlinespace[1.5pt]

\mbox{Qwen3-4B-Instruct}
& GPT-5.1
& \underline{50.98} & \underline{53.60}
& 15.46 & 10.95 \\

GPT-5.1
& GPT-5.1
& \textbf{66.30} & \textbf{58.23}
& 16.24 & 12.08 \\

\bottomrule
\end{tabular}%
}

\vspace{-1.0\baselineskip}
\end{wraptable}

%% file: iclr2027_conference.bib
@article{DBLP:journals/corr/abs-2601-22060,
  author       = {Wenxuan Huang and
                  Yu Zeng and
                  Qiuchen Wang and
                  Zhen Fang and
                  Shaosheng Cao and
                  Zheng Chu and
                  Qingyu Yin and
                  Shuang Chen and
                  Zhenfei Yin and
                  Lin Chen and
                  Zehui Chen and
                  Yao Hu and
                  Philip Torr and
                  Feng Zhao and
                  Wanli Ouyang},
  title        = {Vision-DeepResearch: Incentivizing DeepResearch Capability in Multimodal
                  Large Language Models},
  journal      = {CoRR},
  volume       = {abs/2601.22060},
  year         = {2026},
  url          = {https://doi.org/10.48550/arXiv.2601.22060},
  doi          = {10.48550/ARXIV.2601.22060},
  eprinttype   = {arXiv},
  eprint       = {2601.22060},
  bibsource    = {dblp computer science bibliography, https://dblp.org}
}

@article{DBLP:journals/corr/abs-2601-12346,
  author       = {Peizhou Huang and
                  Zixuan Zhong and
                  Zhongwei Wan and
                  Donghao Zhou and
                  Samiul Alam and
                  Xin Wang and
                  Zexin Li and
                  Zhihao Dou and
                  Li Zhu and
                  Jing Xiong and
                  Chaofan Tao and
                  Yan Xu and
                  Dimitrios Dimitriadis and
                  Tuo Zhang and
                  Mi Zhang},
  title        = {MMDeepResearch-Bench: {A} Benchmark for Multimodal Deep Research Agents},
  journal      = {CoRR},
  volume       = {abs/2601.12346},
  year         = {2026},
  url          = {https://doi.org/10.48550/arXiv.2601.12346},
  doi          = {10.48550/ARXIV.2601.12346},
  eprinttype   = {arXiv},
  eprint       = {2601.12346},
  bibsource    = {dblp computer science bibliography, https://dblp.org}
}

@article{DBLP:journals/corr/abs-2601-09278,
  author       = {Xiaohan Yu and
                  Chao Feng and
                  Lang Mei and
                  Chong Chen},
  title        = {M\({}^{\mbox{3}}\)Searcher: Modular Multimodal Information Seeking
                  Agency with Retrieval-Oriented Reasoning},
  journal      = {CoRR},
  volume       = {abs/2601.09278},
  year         = {2026},
  url          = {https://doi.org/10.48550/arXiv.2601.09278},
  doi          = {10.48550/ARXIV.2601.09278},
  eprinttype   = {arXiv},
  eprint       = {2601.09278},
  bibsource    = {dblp computer science bibliography, https://dblp.org}
}

@article{DBLP:journals/corr/abs-2603-01050,
  author       = {Huanjin Yao and
                  Qixiang Yin and
                  Min Yang and
                  Ziwang Zhao and
                  Yibo Wang and
                  Haotian Luo and
                  Jingyi Zhang and
                  Jiaxing Huang},
  title        = {MM-DeepResearch: {A} Simple and Effective Multimodal Agentic Search
                  Baseline},
  journal      = {CoRR},
  volume       = {abs/2603.01050},
  year         = {2026},
  url          = {https://doi.org/10.48550/arXiv.2603.01050},
  doi          = {10.48550/ARXIV.2603.01050},
  eprinttype   = {arXiv},
  eprint       = {2603.01050},
  bibsource    = {dblp computer science bibliography, https://dblp.org}
}

@misc{zeng2026visiondeepresearchbenchmarkrethinkingvisual,
      title={Vision-DeepResearch Benchmark: Rethinking Visual and Textual Search for Multimodal Large Language Models}, 
      author={Yu Zeng and Wenxuan Huang and Zhen Fang and Shuang Chen and Yufan Shen and Yishuo Cai and Xiaoman Wang and Zhenfei Yin and Lin Chen and Zehui Chen and Shiting Huang and Yiming Zhao and Xu Tang and Yao Hu and Philip Torr and Wanli Ouyang and Shaosheng Cao},
      year={2026},
      eprint={2602.02185},
      archivePrefix={arXiv},
      primaryClass={cs.CV},
      url={https://arxiv.org/abs/2602.02185}, 
}

@misc{huang2026visiondeepresearchincentivizingdeepresearchcapability,
      title={Vision-DeepResearch: Incentivizing DeepResearch Capability in Multimodal Large Language Models}, 
      author={Wenxuan Huang and Yu Zeng and Qiuchen Wang and Zhen Fang and Shaosheng Cao and Zheng Chu and Qingyu Yin and Shuang Chen and Zhenfei Yin and Lin Chen and Zehui Chen and Xu Tang and Yao Hu and Shaohui Lin and Philip Torr and Feng Zhao and Wanli Ouyang},
      year={2026},
      eprint={2601.22060},
      archivePrefix={arXiv},
      primaryClass={cs.CV},
      url={https://arxiv.org/abs/2601.22060}, 
}

@misc{yang2025multimodaldeepresearchergeneratingtextchart,
      title={Multimodal DeepResearcher: Generating Text-Chart Interleaved Reports From Scratch with Agentic Framework}, 
      author={Zhaorui Yang and Bo Pan and Han Wang and Yiyao Wang and Xingyu Liu and Luoxuan Weng and Yingchaojie Feng and Haozhe Feng and Minfeng Zhu and Bo Zhang and Wei Chen},
      year={2025},
      eprint={2506.02454},
      archivePrefix={arXiv},
      primaryClass={cs.CL},
      url={https://arxiv.org/abs/2506.02454}, 
}

@inproceedings{DBLP:conf/acl/XiongYLCD26,
  author       = {Lei Xiong and
                  Huaying Yuan and
                  Zheng Liu and
                  Zhao Cao and
                  Zhicheng Dou},
  editor       = {Maria Liakata and
                  Viviane P. Moreira and
                  Jiajun Zhang and
                  David Jurgens},
  title        = {PaperScope: {A} Multi-Modal Multi-Document Benchmark for Agentic Deep
                  Research Across Massive Scientific Papers},
  booktitle    = {Findings of the Association for Computational Linguistics, {ACL} 2026,
                  San Diego, California, United States, July 2-7, 2026},
  pages        = {8015--8040},
  publisher    = {Association for Computational Linguistics},
  year         = {2026},
  url          = {https://doi.org/10.18653/v1/2026.findings-acl.394},
  doi          = {10.18653/V1/2026.FINDINGS-ACL.394},
  bibsource    = {dblp computer science bibliography, https://dblp.org}
}

@misc{li2026justreasonjustscan,
      title={Not Just Reason, Not Just Scan: Reinforcement Learning for Proactive Scientific Error Verification over Academic Paper}, 
      author={Rongjin Li and Yuanxin Liu and Hao Zhou and Fandong Meng and Jie Zhou and Xu Sun},
      year={2026},
      eprint={2608.26596},
      archivePrefix={arXiv},
      primaryClass={cs.CL},
      url={https://arxiv.org/abs/2608.26596}, 
}

@inproceedings{DBLP:conf/acl/MaZKZCL25,
  author       = {Xueguang Ma and
                  Shengyao Zhuang and
                  Bevan Koopman and
                  Guido Zuccon and
                  Wenhu Chen and
                  Jimmy Lin},
  editor       = {Wanxiang Che and
                  Joyce Nabende and
                  Ekaterina Shutova and
                  Mohammad Taher Pilehvar},
  title        = {{VISA:} Retrieval Augmented Generation with Visual Source Attribution},
  booktitle    = {Proceedings of the 63rd Annual Meeting of the Association for Computational
                  Linguistics (Volume 1: Long Papers), {ACL} 2025, Vienna, Austria,
                  July 27 - August 1, 2025},
  pages        = {30154--30169},
  publisher    = {Association for Computational Linguistics},
  year         = {2025},
  url          = {https://doi.org/10.18653/v1/2025.acl-long.1456},
  doi          = {10.18653/V1/2025.ACL-LONG.1456},
  bibsource    = {dblp computer science bibliography, https://dblp.org}
}

@article{DBLP:journals/corr/abs-2505-19897,
  author       = {Qiushi Sun and
                  Zhoumianze Liu and
                  Chang Ma and
                  Zichen Ding and
                  Fangzhi Xu and
                  Zhangyue Yin and
                  Haiteng Zhao and
                  Zhenyu Wu and
                  Kanzhi Cheng and
                  Zhaoyang Liu and
                  Jianing Wang and
                  Qintong Li and
                  Xiangru Tang and
                  Tianbao Xie and
                  Xiachong Feng and
                  Xiang Li and
                  Ben Kao and
                  Wenhai Wang and
                  Biqing Qi and
                  Lingpeng Kong and
                  Zhiyong Wu},
  title        = {ScienceBoard: Evaluating Multimodal Autonomous Agents in Realistic
                  Scientific Workflows},
  journal      = {CoRR},
  volume       = {abs/2505.19897},
  year         = {2025},
  url          = {https://doi.org/10.48550/arXiv.2505.19897},
  doi          = {10.48550/ARXIV.2505.19897},
  eprinttype   = {arXiv},
  eprint       = {2505.19897},
  bibsource    = {dblp computer science bibliography, https://dblp.org}
}

@inproceedings{DBLP:conf/emnlp/ZhaoJLG25,
  author       = {Suifeng Zhao and
                  Zhuoran Jin and
                  Sujian Li and
                  Jun Gao},
  editor       = {Christos Christodoulopoulos and
                  Tanmoy Chakraborty and
                  Carolyn Rose and
                  Violet Peng},
  title        = {FinRAGBench-V: {A} Benchmark for Multimodal {RAG} with Visual Citation
                  in the Financial Domain},
  booktitle    = {Proceedings of the 2025 Conference on Empirical Methods in Natural
                  Language Processing, {EMNLP} 2025, Suzhou, China, November 4-9, 2025},
  pages        = {4215--4249},
  publisher    = {Association for Computational Linguistics},
  year         = {2025},
  url          = {https://doi.org/10.18653/v1/2025.emnlp-main.211},
  doi          = {10.18653/V1/2025.EMNLP-MAIN.211},
  bibsource    = {dblp computer science bibliography, https://dblp.org}
}

@inproceedings{DBLP:conf/nips/WangDZCCWXHZ25,
  author       = {Qiuchen Wang and
                  Ruixue Ding and
                  Yu Zeng and
                  Zehui Chen and
                  Lin Chen and
                  Shihang Wang and
                  Pengjun Xie and
                  Fei Huang and
                  Feng Zhao},
  editor       = {Danielle Belgrave and
                  Cheng Zhang and
                  Laura N. Montoya and
                  Hsuan{-}Tien Lin and
                  Razvan Pascanu and
                  Piotr Koniusz and
                  Marzyeh Ghassemi and
                  Nancy Chen and
                  Iv{\'{a}}n Vladimir Meza Ru{\'{\i}}z and
                  Arturo Loaiza{-}Bonilla},
  title        = {{VRAG-RL:} Empower Vision-Perception-Based {RAG} for Visually Rich
                  Information Understanding via Iterative Reasoning with Reinforcement
                  Learning},
  booktitle    = {Advances in Neural Information Processing Systems 38: Annual Conference
                  on Neural Information Processing Systems 2025, NeurIPS 2025, San Diego,
                  CA, USA, December 2-7, 2025 / Mexico City, Mexico, November 30 - December
                  5, 2025},
  year         = {2025},
  url          = {http://papers.nips.cc/paper\_files/paper/2025/hash/529b876b5d87f997e98ed925a3ce9f01-Abstract-Conference.html},
  bibsource    = {dblp computer science bibliography, https://dblp.org}
}

@article{DBLP:journals/corr/abs-2602-24134,
  author       = {Zhengren Wang and
                  Dongsheng Ma and
                  Huaping Zhong and
                  Jiayu Li and
                  Wentao Zhang and
                  Bin Wang and
                  Conghui He},
  title        = {AgenticOCR: Parsing Only What You Need for Efficient Retrieval-Augmented
                  Generation},
  journal      = {CoRR},
  volume       = {abs/2602.24134},
  year         = {2026},
  url          = {https://doi.org/10.48550/arXiv.2602.24134},
  doi          = {10.48550/ARXIV.2602.24134},
  eprinttype   = {arXiv},
  eprint       = {2602.24134},
  bibsource    = {dblp computer science bibliography, https://dblp.org}
}

@inproceedings{DBLP:conf/emnlp/WangDCWWXZ25,
  author       = {Qiuchen Wang and
                  Ruixue Ding and
                  Zehui Chen and
                  Weiqi Wu and
                  Shihang Wang and
                  Pengjun Xie and
                  Feng Zhao},
  editor       = {Christos Christodoulopoulos and
                  Tanmoy Chakraborty and
                  Carolyn Rose and
                  Violet Peng},
  title        = {ViDoRAG: Visual Document Retrieval-Augmented Generation via Dynamic
                  Iterative Reasoning Agents},
  booktitle    = {Proceedings of the 2025 Conference on Empirical Methods in Natural
                  Language Processing, {EMNLP} 2025, Suzhou, China, November 4-9, 2025},
  pages        = {9113--9134},
  publisher    = {Association for Computational Linguistics},
  year         = {2025},
  url          = {https://doi.org/10.18653/v1/2025.emnlp-main.464},
  doi          = {10.18653/V1/2025.EMNLP-MAIN.464},
  bibsource    = {dblp computer science bibliography, https://dblp.org}
}

@article{DBLP:journals/corr/abs-2604-14967,
  author       = {Jun Wang and
                  Shuo Tan and
                  Zelong Sun and
                  Tiancheng Gu and
                  Yongle Zhao and
                  Ziyong Feng and
                  Kaicheng Yang and
                  Zhiwu Lu},
  title        = {UniDoc-RL: Coarse-to-Fine Visual {RAG} with Hierarchical Actions and
                  Dense Rewards},
  journal      = {CoRR},
  volume       = {abs/2604.14967},
  year         = {2026},
  url          = {https://doi.org/10.48550/arXiv.2604.14967},
  doi          = {10.48550/ARXIV.2604.14967},
  eprinttype   = {arXiv},
  eprint       = {2604.14967},
  bibsource    = {dblp computer science bibliography, https://dblp.org}
}

@article{shen2026visor,
  title={VISOR: Agentic Visual Retrieval-Augmented Generation via Iterative Search and Over-horizon Reasoning},
  author={Shen, Yucheng and Wu, Jiulong and Huang, Jizhou and Yin, Dawei and Yan, Lingyong and Cao, Min},
  journal={arXiv preprint arXiv:2604.09508},
  year={2026}
}

@article{Zheng2025DeepEyesI,
  title={DeepEyes: Incentivizing "Thinking with Images" via Reinforcement Learning},
  author={Ziwei Zheng and Michael Yang and Jack Hong and Chenxiao Zhao and Guohai Xu and Le Yang and Chao Shen and Xing Yu},
  journal={ArXiv},
  year={2025},
  volume={abs/2505.14362},
  url={https://api.semanticscholar.org/CorpusID:278769859}
}

@misc{gao2026swarmbenchlargelanguagemodels,
      title={SwarmBench: Can Large Language Models Act as Agent Swarm Orchestrators?}, 
      author={Jinshan Gao and Zhuoran Jin and Tianyi Men and Kang Liu and Jun Zhao},
      year={2026},
      eprint={2608.30661},
      archivePrefix={arXiv},
      primaryClass={cs.CL},
      url={https://arxiv.org/abs/2608.30661}, 
}

@misc{phillips2026building,
  author       = {Phillips, Cara},
  title        = {Building multi-agent systems: When and how to use them},
  year         = {2026},
  month        = jan,
  howpublished = {Anthropic Blog},
  url          = {https://claude.com/blog/building-multi-agent-systems-when-and-how-to-use-them}
}

@article{DBLP:journals/corr/abs-2602-04634,
  author       = {Zelai Xu and
                  Zhexuan Xu and
                  Ruize Zhang and
                  Chunyang Zhu and
                  Shi Yu and
                  Weilin Liu and
                  Quanlu Zhang and
                  Wenbo Ding and
                  Chao Yu and
                  Yu Wang},
  title        = {WideSeek-R1: Exploring Width Scaling for Broad Information Seeking
                  via Multi-Agent Reinforcement Learning},
  journal      = {CoRR},
  volume       = {abs/2602.04634},
  year         = {2026},
  url          = {https://doi.org/10.48550/arXiv.2602.04634},
  doi          = {10.48550/ARXIV.2602.04634},
  eprinttype   = {arXiv},
  eprint       = {2602.04634},
  bibsource    = {dblp computer science bibliography, https://dblp.org}
}

@article{DBLP:journals/corr/abs-2607-08662,
  author       = {Xiaoshuai Song and
                  Liancheng Zhang and
                  Kangzhi Zhao and
                  Yutao Zhu and
                  Zhongyuan Wang and
                  Guanting Dong and
                  Jinghan Yang and
                  Han Li and
                  Kun Gai and
                  Ji{-}Rong Wen and
                  Zhicheng Dou},
  title        = {WebSwarm: Recursive Multi-Agent Orchestration for Deep-and-Wide Web
                  Search},
  journal      = {CoRR},
  volume       = {abs/2607.08662},
  year         = {2026},
  url          = {https://doi.org/10.48550/arXiv.2607.08662},
  doi          = {10.48550/ARXIV.2607.08662},
  eprinttype   = {arXiv},
  eprint       = {2607.08662},
  bibsource    = {dblp computer science bibliography, https://dblp.org}
}

@article{DBLP:journals/corr/abs-2602-02276,
  author       = {Kimi Team},
  title        = {Kimi {K2.5:} Visual Agentic Intelligence},
  journal      = {CoRR},
  volume       = {abs/2602.02276},
  year         = {2026},
  url          = {https://doi.org/10.48550/arXiv.2602.02276},
  doi          = {10.48550/ARXIV.2602.02276},
  eprinttype   = {arXiv},
  eprint       = {2602.02276},
  bibsource    = {dblp computer science bibliography, https://dblp.org}
}

@misc{zhang2026orchestrao1omnimodalagentorchestration,
      title={Orchestra-o1: Omnimodal Agent Orchestration}, 
      author={Fan Zhang and Vireo Zhang and Shengju Qian and Haoxuan Li and Hao Wu and Jinyang Wu and Donghao Zhou and Zhihong Zhu and Zheng Lian and Xin Wang and Pheng-Ann Heng},
      year={2026},
      eprint={2606.13707},
      archivePrefix={arXiv},
      primaryClass={cs.AI},
      url={https://arxiv.org/abs/2606.13707}, 
}

@misc{virk2026swarmresearchorchestratingcodingagents,
      title={SwarmResearch: Orchestrating Coding Agents for Open-Ended Discovery}, 
      author={Yuvraj Virk and Zack Edds and Chunqiu Steven Xia and Lingming Zhang},
      year={2026},
      eprint={2607.02807},
      archivePrefix={arXiv},
      primaryClass={cs.AI},
      url={https://arxiv.org/abs/2607.02807}, 
}

@inproceedings{Wu2023AutoGenEN,
  title={AutoGen: Enabling Next-Gen LLM Applications via Multi-Agent Conversation},
  author={Qingyun Wu and Gagan Bansal and Jieyu Zhang and Yiran Wu and Beibin Li and Erkang (Eric) Zhu and Li Jiang and Xiaoyun Zhang and Shaokun Zhang and Jiale Liu and Ahmed Hassan Awadallah and Ryen W. White and Doug Burger and Chi Wang},
  year={2023},
  url={https://api.semanticscholar.org/CorpusID:263611068}
}

@article{Li2023CAMELCA,
  title={CAMEL: Communicative Agents for "Mind" Exploration of Large Language Model Society},
  author={G. Li and Hasan Hammoud and Hani Itani and Dmitrii Khizbullin and Bernard Ghanem},
  journal={Advances in Neural Information Processing Systems 36},
  year={2023},
  url={https://api.semanticscholar.org/CorpusID:268042527}
}

@article{Hong2023MetaGPTMP,
  title={MetaGPT: Meta Programming for Multi-Agent Collaborative Framework},
  author={Sirui Hong and Xiawu Zheng and Jonathan P. Chen and Yuheng Cheng and Ceyao Zhang and Zili Wang and Steven Ka Shing Yau and Zi Hen Lin and Liyang Zhou and Chenyu Ran and Lingfeng Xiao and Chenglin Wu},
  journal={ArXiv},
  year={2023},
  volume={abs/2308.00352},
  url={https://api.semanticscholar.org/CorpusID:260351380}
}

@inproceedings{DBLP:conf/aaai/TanakaNNHSS23,
  author       = {Ryota Tanaka and
                  Kyosuke Nishida and
                  Kosuke Nishida and
                  Taku Hasegawa and
                  Itsumi Saito and
                  Kuniko Saito},
  editor       = {Brian Williams and
                  Yiling Chen and
                  Jennifer Neville},
  title        = {SlideVQA: {A} Dataset for Document Visual Question Answering on Multiple
                  Images},
  booktitle    = {Thirty-Seventh {AAAI} Conference on Artificial Intelligence, {AAAI}
                  2023, Thirty-Fifth Conference on Innovative Applications of Artificial
                  Intelligence, {IAAI} 2023, Thirteenth Symposium on Educational Advances
                  in Artificial Intelligence, {EAAI} 2023, Washington, DC, USA, February
                  7-14, 2023},
  pages        = {13636--13645},
  publisher    = {{AAAI} Press},
  year         = {2023},
  url          = {https://doi.org/10.1609/aaai.v37i11.26598},
  doi          = {10.1609/AAAI.V37I11.26598},
  bibsource    = {dblp computer science bibliography, https://dblp.org}
}

@article{wang2026mineru2,
  title={MinerU2. 5-Pro: Pushing the Limits of Data-Centric Document Parsing at Scale},
  author={Wang, Bin and He, Tianyao and Ouyang, Linke and Wu, Fan and Zhao, Zhiyuan and Chu, Tao and Qu, Yuan and Jin, Zhenjiang and Zeng, Weijun and Miao, Ziyang and others},
  journal={arXiv preprint arXiv:2604.04771},
  year={2026}
}

@article{DBLP:journals/corr/abs-2402-03300,
  author       = {Zhihong Shao and
                  Peiyi Wang and
                  Qihao Zhu and
                  Runxin Xu and
                  Junxiao Song and
                  Mingchuan Zhang and
                  Y. K. Li and
                  Y. Wu and
                  Daya Guo},
  title        = {DeepSeekMath: Pushing the Limits of Mathematical Reasoning in Open
                  Language Models},
  journal      = {CoRR},
  volume       = {abs/2402.03300},
  year         = {2024},
  url          = {https://doi.org/10.48550/arXiv.2402.03300},
  doi          = {10.48550/ARXIV.2402.03300},
  eprinttype    = {arXiv},
  eprint       = {2402.03300},
  bibsource    = {dblp computer science bibliography, https://dblp.org}
}

@article{ma2026citevqa,
  title={CiteVQA: Benchmarking Evidence Attribution for Trustworthy Document Intelligence},
  author={Ma, Dongsheng and Li, Jiayu and Wang, Zhengren and Wang, Yijie and Kong, Jiahao and Zeng, Weijun and Xiao, Jutao and Yang, Jie and Zhang, Wentao and Wang, Bin and others},
  journal={arXiv preprint arXiv:2605.12882},
  year={2026}
}

@article{DBLP:journals/corr/abs-2510-09733,
  author       = {Yubo Sun and
                  Chunyi Peng and
                  Yukun Yan and
                  Shi Yu and
                  Zhenghao Liu and
                  Chi Chen and
                  Zhiyuan Liu and
                  Maosong Sun},
  title        = {VisRAG 2.0: Evidence-Guided Multi-Image Reasoning in Visual Retrieval-Augmented
                  Generation},
  journal      = {CoRR},
  volume       = {abs/2510.09733},
  year         = {2025},
  url          = {https://doi.org/10.48550/arXiv.2510.09733},
  doi          = {10.48550/ARXIV.2510.09733},
  eprinttype    = {arXiv},
  eprint       = {2510.09733},
  bibsource    = {dblp computer science bibliography, https://dblp.org}
}

@misc{yao2023reactsynergizingreasoningacting,
      title={ReAct: Synergizing Reasoning and Acting in Language Models}, 
      author={Shunyu Yao and Jeffrey Zhao and Dian Yu and Nan Du and Izhak Shafran and Karthik Narasimhan and Yuan Cao},
      year={2023},
      eprint={2210.03629},
      archivePrefix={arXiv},
      primaryClass={cs.CL},
      url={https://arxiv.org/abs/2210.03629}, 
}

@article{DBLP:journals/corr/abs-2503-09516,
  author       = {Bowen Jin and
                  Hansi Zeng and
                  Zhenrui Yue and
                  Dong Wang and
                  Hamed Zamani and
                  Jiawei Han},
  title        = {Search-R1: Training LLMs to Reason and Leverage Search Engines with
                  Reinforcement Learning},
  journal      = {CoRR},
  volume       = {abs/2503.09516},
  year         = {2025},
  url          = {https://doi.org/10.48550/arXiv.2503.09516},
  doi          = {10.48550/ARXIV.2503.09516},
  eprinttype    = {arXiv},
  eprint       = {2503.09516},
  bibsource    = {dblp computer science bibliography, https://dblp.org}
}

@article{DBLP:journals/corr/abs-2601-06021,
  author       = {Jiajie Zhang and
                  Xin Lv and
                  Ling Feng and
                  Lei Hou and
                  Juanzi Li},
  title        = {Chaining the Evidence: Robust Reinforcement Learning for Deep Search
                  Agents with Citation-Aware Rubric Rewards},
  journal      = {CoRR},
  volume       = {abs/2601.06021},
  year         = {2026},
  url          = {https://doi.org/10.48550/arXiv.2601.06021},
  doi          = {10.48550/ARXIV.2601.06021},
  eprinttype    = {arXiv},
  eprint       = {2601.06021},
  bibsource    = {dblp computer science bibliography, https://dblp.org}
}

@article{DBLP:journals/corr/abs-2601-04525,
  author       = {Yibo Zhao and
                  Jiapeng Zhu and
                  Zichen Ding and
                  Xiang Li},
  title        = {{GRACE:} Reinforcement Learning for Grounded Response and Abstention
                  under Contextual Evidence},
  journal      = {CoRR},
  volume       = {abs/2601.04525},
  year         = {2026},
  url          = {https://doi.org/10.48550/arXiv.2601.04525},
  doi          = {10.48550/ARXIV.2601.04525},
  eprinttype    = {arXiv},
  eprint       = {2601.04525},
  bibsource    = {dblp computer science bibliography, https://dblp.org}
}

@misc{zhao2026mciterlreliablemultimodalrag,
      title={MCite-RL: Towards Reliable Multimodal RAG via Citation-enhanced Agentic Reinforcement Learning}, 
      author={Suifeng Zhao and Zida Liu and Xinyu Lei and Lei Sun and Jun Gao and Sujian Li},
      year={2026},
      eprint={2608.21808},
      archivePrefix={arXiv},
      primaryClass={cs.CL},
      url={https://arxiv.org/abs/2608.21808}, 
}

@misc{qwen3technicalreport,
      title={Qwen3 Technical Report}, 
      author={Qwen Team},
      year={2025},
      eprint={2505.09388},
      archivePrefix={arXiv},
      primaryClass={cs.CL},
      url={https://arxiv.org/abs/2505.09388}, 
}

@misc{singh2026openaigpt5card,
      title={OpenAI GPT-5 System Card}, 
      author={Aaditya Singh and Adam Fry and Adam Perelman and Adam Tart and Adi Ganesh and Ahmed El-Kishky and Aidan McLaughlin and Aiden Low and AJ Ostrow and Akhila Ananthram and Akshay Nathan and Alan Luo and Alec Helyar and Aleksander Madry and Aleksandr Efremov and Aleksandra Spyra and Alex Baker-Whitcomb and Alex Beutel and Alex Karpenko and Alex Makelov and Alex Neitz and Alex Wei and Alexandra Barr and Alexandre Kirchmeyer and Alexey Ivanov and Alexi Christakis and Alistair Gillespie and Allison Tam and Ally Bennett and Alvin Wan and Alyssa Huang and Amy McDonald Sandjideh and Amy Yang and Ananya Kumar and Andre Saraiva and Andrea Vallone and Andrei Gheorghe and Andres Garcia Garcia and Andrew Braunstein and Andrew Liu and Andrew Schmidt and Andrey Mereskin and Andrey Mishchenko and Andy Applebaum and Andy Rogerson and Ann Rajan and Annie Wei and Anoop Kotha and Anubha Srivastava and Anushree Agrawal and Arun Vijayvergiya and Ashley Tyra and Ashvin Nair and Avi Nayak and Ben Eggers and Bessie Ji and Beth Hoover and Bill Chen and Blair Chen and Boaz Barak and Borys Minaiev and Botao Hao and Bowen Baker and Brad Lightcap and Brandon McKinzie and Brandon Wang and Brendan Quinn and Brian Fioca and Brian Hsu and Brian Yang and Brian Yu and Brian Zhang and Brittany Brenner and Callie Riggins Zetino and Cameron Raymond and Camillo Lugaresi and Carolina Paz and Cary Hudson and Cedric Whitney and Chak Li and Charles Chen and Charlotte Cole and Chelsea Voss and Chen Ding and Chen Shen and Chengdu Huang and Chris Colby and Chris Hallacy and Chris Koch and Chris Lu and Christina Kaplan and Christina Kim and CJ Minott-Henriques and Cliff Frey and Cody Yu and Coley Czarnecki and Colin Reid and Colin Wei and Cory Decareaux and Cristina Scheau and Cyril Zhang and Cyrus Forbes and Da Tang and Dakota Goldberg and Dan Roberts and Dana Palmie and Daniel Kappler and Daniel Levine and Daniel Wright and Dave Leo and David Lin and David Robinson and Declan Grabb and Derek Chen and Derek Lim and Derek Salama and Dibya Bhattacharjee and Dimitris Tsipras and Dinghua Li and Dingli Yu and DJ Strouse and Drew Williams and Dylan Hunn and Ed Bayes and Edwin Arbus and Ekin Akyurek and Elaine Ya Le and Elana Widmann and Eli Yani and Elizabeth Proehl and Enis Sert and Enoch Cheung and Eri Schwartz and Eric Han and Eric Jiang and Eric Mitchell and Eric Sigler and Eric Wallace and Erik Ritter and Erin Kavanaugh and Evan Mays and Evgenii Nikishin and Fangyuan Li and Felipe Petroski Such and Filipe de Avila Belbute Peres and Filippo Raso and Florent Bekerman and Foivos Tsimpourlas and Fotis Chantzis and Francis Song and Francis Zhang and Gaby Raila and Garrett McGrath and Gary Briggs and Gary Yang and Giambattista Parascandolo and Gildas Chabot and Grace Kim and Grace Zhao and Gregory Valiant and Guillaume Leclerc and Hadi Salman and Hanson Wang and Hao Sheng and Haoming Jiang and Haoyu Wang and Haozhun Jin and Harshit Sikchi and Heather Schmidt and Henry Aspegren and Honglin Chen and Huida Qiu and Hunter Lightman and Ian Covert and Ian Kivlichan and Ian Silber and Ian Sohl and Ibrahim Hammoud and Ignasi Clavera and Ikai Lan and Ilge Akkaya and Ilya Kostrikov and Irina Kofman and Isak Etinger and Ishaan Singal and Jackie Hehir and Jacob Huh and Jacqueline Pan and Jake Wilczynski and Jakub Pachocki and James Lee and James Quinn and Jamie Kiros and Janvi Kalra and Jasmyn Samaroo and Jason Wang and Jason Wolfe and Jay Chen and Jay Wang and Jean Harb and Jeffrey Han and Jeffrey Wang and Jennifer Zhao and Jeremy Chen and Jerene Yang and Jerry Tworek and Jesse Chand and Jessica Landon and Jessica Liang and Ji Lin and Jiancheng Liu and Jianfeng Wang and Jie Tang and Jihan Yin and Joanne Jang and Joel Morris and Joey Flynn and Johannes Ferstad and Johannes Heidecke and John Fishbein and John Hallman and Jonah Grant and Jonathan Chien and Jonathan Gordon and Jongsoo Park and Jordan Liss and Jos Kraaijeveld and Joseph Guay and Joseph Mo and Josh Lawson and Josh McGrath and Joshua Vendrow and Joy Jiao and Julian Lee and Julie Steele and Julie Wang and Junhua Mao and Kai Chen and Kai Hayashi and Kai Xiao and Kamyar Salahi and Kan Wu and Karan Sekhri and Karan Sharma and Karan Singhal and Karen Li and Kenny Nguyen and Keren Gu-Lemberg and Kevin King and Kevin Liu and Kevin Stone and Kevin Yu and Kristen Ying and Kristian Georgiev and Kristie Lim and Kushal Tirumala and Kyle Miller and Lama Ahmad and Larry Lv and Laura Clare and Laurance Fauconnet and Lauren Itow and Lauren Yang and Laurentia Romaniuk and Leah Anise and Lee Byron and Leher Pathak and Leon Maksin and Leyan Lo and Leyton Ho and Li Jing and Liang Wu and Liang Xiong and Lien Mamitsuka and Lin Yang and Lindsay McCallum and Lindsey Held and Liz Bourgeois and Logan Engstrom and Lorenz Kuhn and Louis Feuvrier and Lu Zhang and Lucas Switzer and Lukas Kondraciuk and Lukasz Kaiser and Manas Joglekar and Mandeep Singh and Mandip Shah and Manuka Stratta and Marcus Williams and Mark Chen and Mark Sun and Marselus Cayton and Martin Li and Marvin Zhang and Marwan Aljubeh and Matt Nichols and Matthew Haines and Max Schwarzer and Mayank Gupta and Meghan Shah and Melody Y. Guan and Melody Huang and Meng Dong and Mengqing Wang and Mia Glaese and Micah Carroll and Michael Lampe and Michael Malek and Michael Sharman and Michael Zhang and Michele Wang and Michelle Pokrass and Mihai Florian and Mikhail Pavlov and Miles Wang and Ming Chen and Mingxuan Wang and Minnia Feng and Mo Bavarian and Molly Lin and Moose Abdool and Mostafa Rohaninejad and Nacho Soto and Natalie Staudacher and Natan LaFontaine and Nathan Marwell and Nelson Liu and Nick Preston and Nick Turley and Nicklas Ansman and Nicole Blades and Nikil Pancha and Nikita Mikhaylin and Niko Felix and Nikunj Handa and Nishant Rai and Nitish Keskar and Noam Brown and Ofir Nachum and Oleg Boiko and Oleg Murk and Olivia Watkins and Oona Gleeson and Pamela Mishkin and Patryk Lesiewicz and Paul Baltescu and Pavel Belov and Peter Zhokhov and Philip Pronin and Phillip Guo and Phoebe Thacker and Qi Liu and Qiming Yuan and Qinghua Liu and Rachel Dias and Rachel Puckett and Rahul Arora and Ravi Teja Mullapudi and Raz Gaon and Reah Miyara and Rennie Song and Rishabh Aggarwal and RJ Marsan and Robel Yemiru and Robert Xiong and Rohan Kshirsagar and Rohan Nuttall and Roman Tsiupa and Ronen Eldan and Rose Wang and Roshan James and Roy Ziv and Rui Shu and Ruslan Nigmatullin and Saachi Jain and Saam Talaie and Sam Altman and Sam Arnesen and Sam Toizer and Sam Toyer and Samuel Miserendino and Sandhini Agarwal and Sarah Yoo and Savannah Heon and Scott Ethersmith and Sean Grove and Sean Taylor and Sebastien Bubeck and Sever Banesiu and Shaokyi Amdo and Shengjia Zhao and Sherwin Wu and Shibani Santurkar and Shiyu Zhao and Shraman Ray Chaudhuri and Shreyas Krishnaswamy and Shuaiqi and Xia and Shuyang Cheng and Shyamal Anadkat and Simón Posada Fishman and Simon Tobin and Siyuan Fu and Somay Jain and Song Mei and Sonya Egoian and Spencer Kim and Spug Golden and SQ Mah and Steph Lin and Stephen Imm and Steve Sharpe and Steve Yadlowsky and Sulman Choudhry and Sungwon Eum and Suvansh Sanjeev and Tabarak Khan and Tal Stramer and Tao Wang and Tao Xin and Tarun Gogineni and Taya Christianson and Ted Sanders and Tejal Patwardhan and Thomas Degry and Thomas Shadwell and Tianfu Fu and Tianshi Gao and Timur Garipov and Tina Sriskandarajah and Toki Sherbakov and Tomek Korbak and Tomer Kaftan and Tomo Hiratsuka and Tongzhou Wang and Tony Song and Tony Zhao and Troy Peterson and Val Kharitonov and Victoria Chernova and Vineet Kosaraju and Vishal Kuo and Vitchyr Pong and Vivek Verma and Vlad Petrov and Wanning Jiang and Weixing Zhang and Wenda Zhou and Wenlei Xie and Wenting Zhan and Wes McCabe and Will DePue and Will Ellsworth and Wulfie Bain and Wyatt Thompson and Xiangning Chen and Xiangyu Qi and Xin Xiang and Xinwei Shi and Yann Dubois and Yaodong Yu and Yara Khakbaz and Yifan Wu and Yilei Qian and Yin Tat Lee and Yinbo Chen and Yizhen Zhang and Yizhong Xiong and Yonglong Tian and Young Cha and Yu Bai and Yu Yang and Yuan Yuan and Yuanzhi Li and Yufeng Zhang and Yuguang Yang and Yujia Jin and Yun Jiang and Yunyun Wang and Yushi Wang and Yutian Liu and Zach Stubenvoll and Zehao Dou and Zheng Wu and Zhigang Wang},
      year={2026},
      eprint={2601.03267},
      archivePrefix={arXiv},
      primaryClass={cs.CL},
      url={https://arxiv.org/abs/2601.03267}, 
}

@article{DBLP:journals/corr/abs-2507-06261,
  author       = {Gheorghe Comanici and
                  Eric Bieber and
                  Mike Schaekermann and
                  Ice Pasupat and
                  Noveen Sachdeva and
                  Inderjit S. Dhillon and
                  Marcel Blistein and
                  Ori Ram and
                  Dan Zhang and
                  Evan Rosen and
                  Luke Marris and
                  Sam Petulla and
                  Colin Gaffney and
                  Asaf Aharoni and
                  Nathan Lintz and
                  Tiago Cardal Pais and
                  Henrik Jacobsson and
                  Idan Szpektor and
                  Nan{-}Jiang Jiang and
                  Krishna Haridasan and
                  Ahmed Omran and
                  Nikunj Saunshi and
                  Dara Bahri and
                  Gaurav Mishra and
                  Eric Chu and
                  Toby Boyd and
                  Brad Hekman and
                  Aaron Parisi and
                  Chaoyi Zhang and
                  Kornraphop Kawintiranon and
                  Tania Bedrax{-}Weiss and
                  Oliver Wang and
                  Ya Xu and
                  Ollie Purkiss and
                  Uri Mendlovic and
                  Ila{\"{\i}} Deutel and
                  Nam Nguyen and
                  Adam Langley and
                  Flip Korn and
                  Lucia Rossazza and
                  Alexandre Ram{\'{e}} and
                  Sagar Waghmare and
                  Helen Miller and
                  Nathan Byrd and
                  Ashrith Sheshan and
                  Raia Hadsell Sangnie Bhardwaj and
                  Pawel Janus and
                  Tero Rissa and
                  Dan Horgan and
                  Sharon Silver and
                  Ayzaan Wahid and
                  Sergey Brin and
                  Yves Raimond and
                  Klemen Kloboves and
                  Cindy Wang and
                  Nitesh Bharadwaj Gundavarapu and
                  Ilia Shumailov and
                  Bo Wang and
                  Mantas Pajarskas and
                  Joe Heyward and
                  Martin Nikoltchev and
                  Maciej Kula and
                  Hao Zhou and
                  Zachary Garrett and
                  Sushant Kafle and
                  Sercan Arik and
                  Ankita Goel and
                  Mingyao Yang and
                  Jiho Park and
                  Koji Kojima and
                  Parsa Mahmoudieh and
                  Koray Kavukcuoglu and
                  Grace Chen and
                  Doug Fritz and
                  Anton Bulyenov and
                  Sudeshna Roy and
                  Dimitris Paparas and
                  Hadar Shemtov and
                  Bo{-}Juen Chen and
                  Robin Strudel and
                  David Reitter and
                  Aurko Roy and
                  Andrey Vlasov and
                  Changwan Ryu and
                  Chas Leichner and
                  Haichuan Yang and
                  Zelda Mariet and
                  Denis Vnukov and
                  Tim Sohn and
                  Amy Stuart and
                  Wei Liang and
                  Minmin Chen and
                  Praynaa Rawlani and
                  Christy Koh and
                  JD Co{-}Reyes and
                  Guangda Lai and
                  Praseem Banzal and
                  Dimitrios Vytiniotis and
                  Jieru Mei and
                  Mu Cai},
  title        = {Gemini 2.5: Pushing the Frontier with Advanced Reasoning, Multimodality,
                  Long Context, and Next Generation Agentic Capabilities},
  journal      = {CoRR},
  volume       = {abs/2507.06261},
  year         = {2025},
  url          = {https://doi.org/10.48550/arXiv.2507.06261},
  doi          = {10.48550/ARXIV.2507.06261},
  eprinttype    = {arXiv},
  eprint       = {2507.06261},
  bibsource    = {dblp computer science bibliography, https://dblp.org}
}

@misc{anthropic2024claude35sonnet,
  author       = {Anthropic},
  title        = {Claude 3.5 Sonnet Raises the Industry Bar for Intelligence},
  howpublished = {\url{https://www.anthropic.com/news/claude-3-5-sonnet}},
  note         = {Accessed: 2025-09-07},
  year         = {2024},
  month        = {June}
}

@misc{faysse2024colpaliefficientdocumentretrieval,
  title={ColPali: Efficient Document Retrieval with Vision Language Models}, 
  author={Manuel Faysse and Hugues Sibille and Tony Wu and Bilel Omrani and Gautier Viaud and Céline Hudelot and Pierre Colombo},
  year={2024},
  eprint={2407.01449},
  archivePrefix={arXiv},
  primaryClass={cs.IR},
  url={https://arxiv.org/abs/2407.01449}, 
}
